\documentclass[lettersize,journal]{IEEEtran}
\usepackage{ulem}
\usepackage{amsmath,amsfonts}
\usepackage{algorithmic}
\usepackage{algorithm}
\usepackage{array}
\usepackage{bm}
\usepackage{textcomp}
\usepackage{stfloats}
\usepackage{url}
\usepackage{verbatim}
\usepackage{graphicx}
\usepackage{subfigure}
\usepackage{caption}
\usepackage{subcaption}
\usepackage{multirow}
\usepackage{booktabs}
\usepackage{cite}
\usepackage{xcolor}
\usepackage{makecell}

\begin{document}

\title{SF-Speech: Straightened Flow for Zero-Shot Voice Clone}

\author{Xuyuan Li\IEEEauthorrefmark{2}\IEEEauthorrefmark{4}, Zengqiang Shang\IEEEauthorrefmark{1}\IEEEauthorrefmark{2}, Hua Hua\IEEEauthorrefmark{2}\IEEEauthorrefmark{4}, Peiyang Shi\IEEEauthorrefmark{2}, Chen Yang\IEEEauthorrefmark{2}\IEEEauthorrefmark{4}, \\Li Wang\IEEEauthorrefmark{2}, and Pengyuan Zhang\thanks{* Corresponding author.}\IEEEauthorrefmark{1}\IEEEauthorrefmark{2}\IEEEauthorrefmark{4} \\
\IEEEauthorblockA{\IEEEauthorrefmark{2}Laboratory of Speech and Intelligent Information Processing, Institute of Acoustics, CAS, China \\ Email: \texttt{\{lixuyuan, shangzengqiang, huahua, shipeiyang, yangchen, wangli, zhangpengyuan\}@hccl.ioa.ac.cn}\\
\IEEEauthorrefmark{4}University of Chinese Academy of Sciences, China\\}
}

\markboth{Journal of \LaTeX\ Class Files,~Vol.~14, No.~8, August~2021}%
{Shell \MakeLowercase{\textit{et al.}}: A Sample Article Using IEEEtran.cls for IEEE Journals}

\IEEEpubid{0000--0000/00\$00.00~\copyright~2021 IEEE}

\maketitle

\begin{abstract}
Recently, neural ordinary differential equations (ODE) models trained with flow matching have achieved impressive performance on the zero-shot voice clone task. Nevertheless, postulating standard Gaussian noise as the initial distribution of ODE gives rise to numerous intersections within the fitted targets of flow matching, which presents challenges to model training and enhances the curvature of the learned generated trajectories. These curved trajectories restrict the capacity of ODE models for generating desirable samples with a few steps. This paper proposes SF-Speech, a novel voice clone model based on ODE and in-context learning. Unlike the previous works, SF-Speech adopts a lightweight multi-stage module to generate a more deterministic initial distribution for ODE. Without introducing any additional loss function, we effectively straighten the curved reverse trajectories of the ODE model by jointly training it with the proposed module. Experiment results on datasets of various scales show that SF-Speech outperforms the state-of-the-art zero-shot TTS methods and requires only a quarter of the solver steps, resulting in a generation speed approximately 3.7 times that of Voicebox and E2 TTS. Audio samples are available at the demo page\footnote{[Online] Available: https://lixuyuan102.github.io/Demo/}.

\end{abstract}




\section{Introduction}

\IEEEPARstart{M}{u}lti-speaker speech generation has been able to synthesize speech similar to human quality, benefiting from the development of neural networks. This technology has been widely applied in voice assistants, audiobooks, video dubbing, etc. As an essential branch of multi-speaker speech generation, cloning the unseen voice has attracted extensive attention. Some earlier works, like \cite{gibiansky2017deep, moss2020boffin, huang2022meta}, have approached this goal though exploring more efficient fine-tuning methods with speech from unseen speakers. However, they are difficult to be effective when the unseen speaker data is only a few seconds long. To address this issue, some works 
\cite{min2021meta, jiang2023mega, casanova2022yourtts,li2023freevc} maintain a hidden space containing the global speaker embedding (GSE) extracted from the reference speech. As a result, the model trained with a large number of samples from this space can deal with the GSE of unseen speech in this hidden space directly. Nevertheless, the GSE with the shape of 1×N is a bottleneck feature where the speaker information is highly compressed, making it more challenging to reconstruct the prosody and timbre of unseen speakers based on this embedding.


In recent years, large-scale speech generation models have shown outstanding performance on the zero-shot voice clone task, leveraging in-context learning and large-scale datasets. For instance, Wang et al. \cite{wang2023neural, yang2023uniaudio, casanova2024xtts} employ large language model (LLM) \cite{floridi2020gpt, touvron2023llama} autoregressively continue the reference speech in discrete space constructed by audio codec models \cite{zeghidour2021soundstream, defossez2022high, kumar2024high}. NaturalSpeech3 \cite{ju2024naturalspeech} constructs a discrete codec model designed to decouple speech components and continue different speech components non-autoregressively with discrete diffusion models \cite{austin2021structured}. Although these discrete token-based methods generated impressive natural and diverse speech, they rely on massive sufficient training data and yield unsatisfactory results on small-scale training data \cite{lajszczak2024base}. Moreover, some studies \cite{meng2024autoregressive,wang2024evaluating,qiang2024high,shen2023naturalspeech} show that discrete tokens exhibit lower audio reconstruction quality compared to continuous acoustic features. In addition to the approaches that model in discrete space, some works apply in-context learning directly to the mel-spectrogram of speech. Le et al. \cite{le2023voicebox, kim2024p} trained a neural ordinary differential equation (ODE) model \cite{chen2018neural} with flow matching (FM) \cite{lipman2022flow} to reconstruct the masked mel-spectrograms from random noise. Building upon these studies, Eskimez et al. \cite{eskimez2024e2, chen2024f5, yang2024simplespeech} incorporated implicit duration modeling. This innovative approach enabled them to avoid complex data processing as well as duration model design and achieve a new state-of-the-art (SOTA) performance.

\IEEEpubidadjcol

While ODE-based approaches have demonstrated remarkable success, they demand numerous evaluations of expensive neural networks during iterative numerical simulations, leading to slow inference. This limitation stems from the intrinsic relationship between the reverse trajectory curvature of ODE and numerical solver efficiency: increased trajectory curvature directly amplifies truncation errors, necessitating a greater number of function evaluations (NFEs) to maintain solution accuracy. In Sec. \ref{OMFM}, we reveal that the reverse trajectory curvature of the ODE-based model trained with FM will become larger as the intersections between forward trajectories increases. Moreover, these crossover points also introduce significant challenges for neural networks fitting forward trajectories during FM training. Previous works \cite{lee2023minimizing, liu2022flow} have found that these intersections demonstrate a positive correlation with the indeterminacy in the coupling between the initial and final distributions of ODE. This insight naturally leads us to examine prevailing initialization strategies in existing works. Conventionally, the approaches that apply the ODE model to speech generation employ standard Gaussian noise as the initial distribution. This practice inherently creates an uncertain mapping between a target mel spectrogram and any noise instances during training, inducing excessive intersection between forward trajectories. Consequently, the reverse generation trajectories learned by these methods may not match the real forward trajectories accurately and require a great NFEs to simulate it.

To address this limitation, we present SF-Speech, a zero-shot voice cloning framework that delivers high-quality synthesis and accelerated inference. Building on the conditional generation paradigm of previous ODE-based systems \cite{le2023voicebox, kim2024p} for in-context audio-guided speech synthesis, our key innovation lies in designing a stable initial distribution for the neural ODE model. Specifically, SF-Speech is a multi-stage model. It first generates a coarse feature, and then uses a neural ODE module to map this feature to the speech mel-spectrogram. To obtain this coarse feature, we design a lightweight two-stage module (7M parameters) to progressively incorporate text-related information and speaker-related information. Since this information represents the main components of speech, the coarse feature can establish a deterministic mapping to the target mel spectrogram. It effectively mitigates forward trajectory intersections in FM, consequently reducing modeling complexity and straightening the learned reverse trajectory. In addition, we enhance the transformer-based ODE backbone from VoiceBox \cite{le2023voicebox} with 2D convolutions, which allows for better reconstruction of local details in the generated mel spectrogram.

We compare SF-Speech with leading zero-shot voice clone models on two datasets: 1) Emilia, a TTS-oriented dataset containing around 100k high-quality speech across Chinese and English. 2) MagicData, a more in-the-wild corpus with 755 hours of Chinese speech. Experimental results on the large-scale dataset demonstrate that SF-Speech achieves new SOTA performance with faster inference speed. When tested on MagicData, the advantages of SF-Speech become even more conspicuous. Across multiple metrics, it either outperforms or is comparable with baseline models that have more parameters or are trained on larger datasets. Furthermore, to investigate the impact of the proposed method on the FM-based ODE model, we measured the curvature of the reverse trajectories for different ODE-based TTS models and then analyzed how the trajectory curvature affects speech generation in depth.

In summary, the main contributions of this paper include:

\begin{itemize}

\item We present SF-Speech, a novel TTS framework that regards speech generation as a three-stage process. This process, based on text-related features, gradually integrates speaker-related information and spectrogram details, leading to more natural and personalized speech synthesis.

\item We design a stable and reliable initial distribution for the FM-based ODE module. This initial distribution reduces the uncertainty in FM training, decreasing the modeling complexity and straightening the reverse trajectories learned by the network.

\item Through systematic experiments on datasets of varying scales and quality, we demonstrate that the proposed approach enhances the performance of ODE-based TTS, with analyses revealing the reason for this improvement.



\end{itemize}


The rest of this paper is structured as follows. In Sec. \ref{rw}, we introduce the related works on differential equations model and large-scale zero-shot voice clone methods. Sec. \ref{sf} delves into the theoretical foundation of the proposed method and then presents the relevant model design. The implementation details, baselines, and evaluation metrics are provided in Sec. \ref{ee}. In Sec. \ref{er} we provide the experimental results and conduct in-depth analyses of them. Finally, Sec. \ref{cl} summarizes our contributions and discusses potential future research directions.

\section{Related work}
\label{rw}

{\bf{Differential equation model:}} From the point of view of differential equations, diffusion models on continuous space can be categorized into stochastic differential equation (SDE) and ODE models. Recent years, the SDE model \cite{song2020score} trained with denoise score matching \cite{hyvarinen2005estimation} have achieved stunning results in image generation \cite{rombach2022high,ramesh2022hierarchical}, vocoders for speech \cite{kong2020diffwave,chen2020wavegrad,koizumi2022specgrad}, and acoustic modeling \cite{popov2021grad,jing2023u}. However, it usually requires hundreds of iterations to generate acceptable samples. As a result, a great deal of follow-up studies have focused on reducing the inference iterations by enhancing the express capability of diffusion noise \cite{zheng2022truncated,lyu2022accelerating,kong2021fast} or designing a new reverse solver \cite{lu2022dpm, sun2017ito}. 

Different from the SDE model, conventional training of ODE models requires complex ODE solvers \cite{chen2018neural}. Lipman et al. \cite{lipman2022flow} proposed a simpler training method, called FM, which allows the ODE model to generate satisfactory samples within a few inference steps without adjusting the diffusion noise or reverse solver. Recently, Le et al. \cite{le2023voicebox, kim2024p, mehta2024matcha} introduced the ODE model trained with this approach to speech generation and achieved results that outperform those of SDE-based models. However, some studies \cite{liu2022flow, lee2023minimizing} found that sampling initial points from a standard normal distribution produces a large number of intersections in the forward trajectory of FM. When such intersecting forward trajectories are employed as the fitting target in training, the network faces a case where one input matches different targets and generates curved reverse trajectories. To address this problem, Liu et al. \cite{liu2022flow} introduced multiple rectified flows to turn the arbitrary coupling between initial and target distribution into a new deterministic coupling. Guo et al. \cite{guo2024voiceflow} introduced this into speech generation. However, the cumulative error due to multiple rectifying significantly impairs the inference performance, especially as the inference steps increase. Lee et al. \cite{lee2023minimizing} introduced another method that reduces the trajectory curvature in one training. They employ a $\beta$-VAE \cite{higgins2016beta} to get a latent distribution that explicitly corresponds to each attribute of images as the initial distribution. In contrast to their work, this paper presents an innovative approach to obtain the initial distribution that couples with speech deterministically in real sample space, without adding additional loss.

{\bf{Large-scale voice clone model:}} Traditional zero-shot voice cloning models exhibit a significant performance gap in speech synthesis for unseen and seen speakers, limited by the representation capacity of GSE and insufficient data. Recently, large-scale speech synthesis models have effectively mitigated this gap, benefiting from in-context learning and large-scale speech datasets. These models can be categorized into three types according to their inference forms: 1) Autoregressive (AR) models, 2) Non-autoregressive (NAR) models, and 3) Hybrid AR + NAR models. VALL-E \cite{wang2023neural} is the first AR large-scale speech synthesis model, which quantizes speech into discrete token sequences via Encodec \cite{defossez2022high} and predicts them autoregressively using a neural language model. Building on this, Uniaudio \cite{yang2023uniaudio} extended the framework to unify 11 audio generation tasks. However, instability generation and slow inference brought by the AR models have endured over time. To address these issues, XTTS \cite{casanova2024xtts} enhanced timbre stability through speaker-consistency loss, while VALL-E2 \cite{chen2024vall} improved sampling stability and reduced sequence length via repetition aware sampling and grouped code modeling.

On the other hand, diffusion techniques have been widely adopted in NAR methods. For instance, NaturalSpeech2 \cite{shen2023naturalspeech} and 3 \cite{ju2024naturalspeech} employed diffusion models in latent continuous and discrete spaces, respectively. Similarly, VoiceBox \cite{le2023voicebox} and P-Flow \cite{kim2024p} applied FM-based neural ODE models to mel-spectrograms. Although these NAR methods outperform AR methods in terms of timbre similarity and stability, they require aligned text-audio pairs for training. To avoid this data preparation, some works \cite{anastassiou2024seed,du2024cosyvoice,du2024cosyvoice2} used language models to generate semantic features aligned with speech and then employed NAR models to convert them into acoustic features. Among these approaches, the ODE model trained with FM remains a crucial component for improving speech quality and similarity. There are also some concurrent works, such as SimpleSpeech2 \cite{yang2024simplespeech}, E2 TTS \cite{eskimez2024e2}, and F5-TTS \cite{chen2024f5}, which explore implicit modeling of the alignment between content and generated speech using only ODE models and achieve new SOTA performance. However, they still follow the initial distribution assumptions in the original paper \cite{lipman2022flow} of FM, which has been proven to limit the performance of ODE models, as we discussed above.


\begin{figure} [h]

\includegraphics[width=\linewidth]{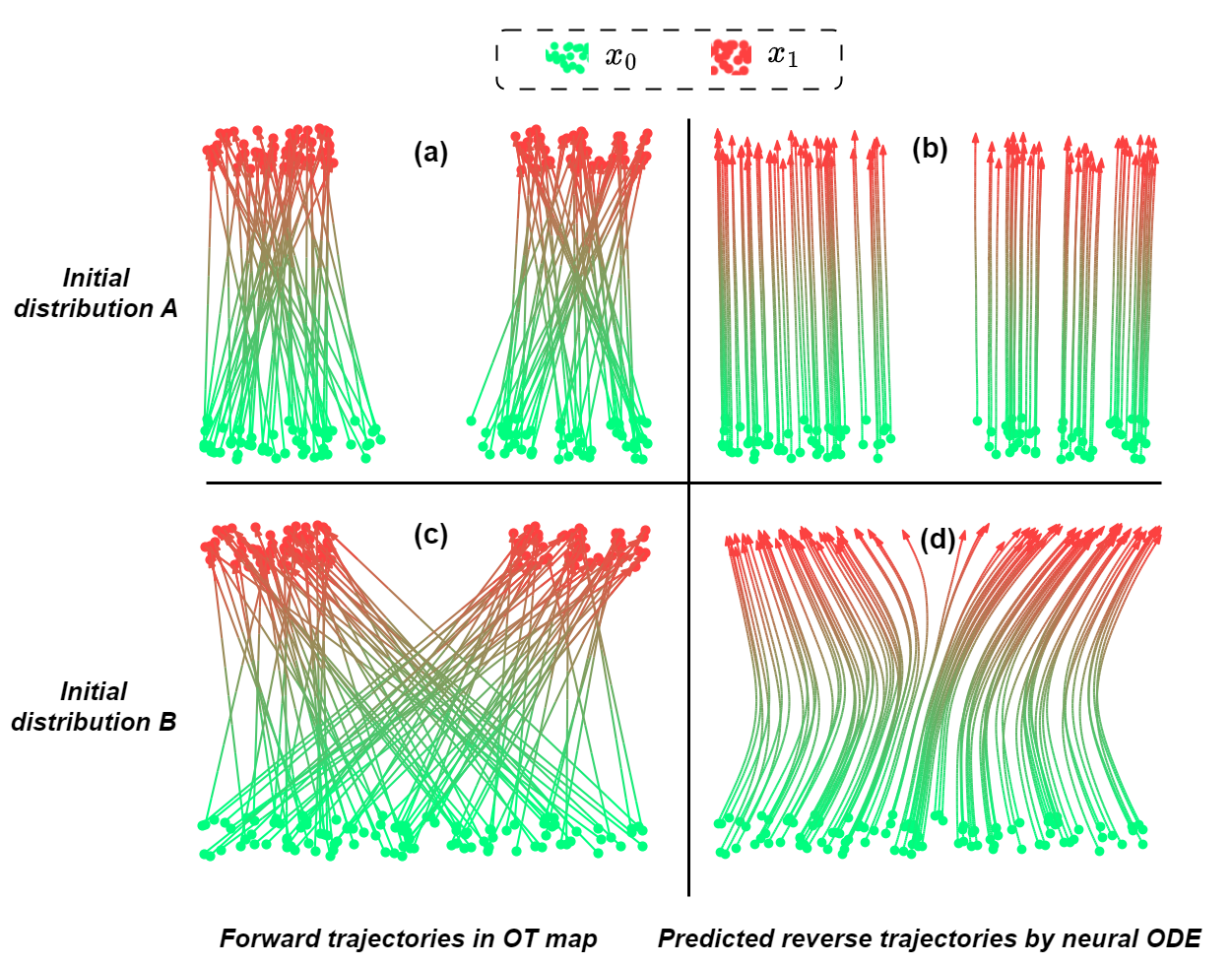}

\centering
\caption{Non-causal direction of forward trajectories((a),(c)) $\&$ Causal direction of reverse trajectories((b),(d)) predicted by the neural network estimator with 128 reverse steps. The upper((a),(b)) and lower((c),(d)) parts show the effect of different initial distributions on learned reverse trajectories.}
\label{fig1}
\end{figure}

\begin{figure*} [h]
        \begin{minipage}{0.90\linewidth}
            \centerline{\includegraphics[width=\textwidth]{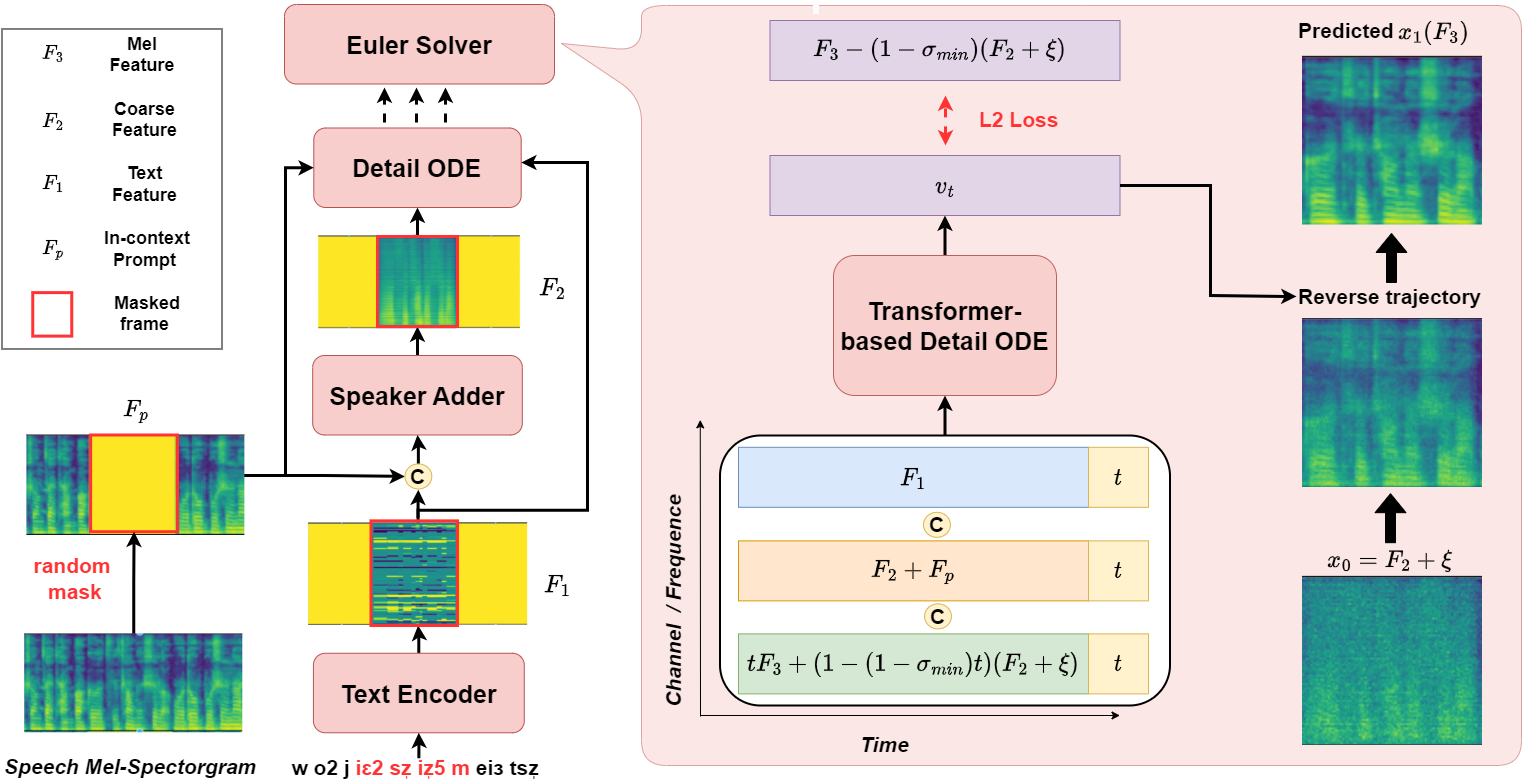}}
        \end{minipage}
\centering
\caption{Overall architecture of SF-Speech (left) and working diagram of ODE model (right) in SF-Speech.}
\label{fig2}
\end{figure*}

\section{SF-Speech} 
\label{sf}
This part presents the theoretical foundation and architecture of the proposed SF-Speech. As shown in Fig. \ref{fig2}, SF-Speech is a model built on neural ODE. Unlike conventional methods, it innovatively introduces a new initial distribution to mitigate the inherent limitations of neural ODE trained with FM, thus improving the quality of the generated speech, as well as the inference efficiency. In Sec. \ref{OMFM}, we first analyze the limitations of the ODE model trained using FM from a theoretical perspective and present a potential solution through experiments on the point transmission task. Subsequently, Sec. \ref{OM} illustrates how SF-Speech applies this solution to zero-shot voice cloning. Finally, we introduce each module of SF-Speech in Sec. \ref{af}, \ref{cf}, and \ref{sss} one by one.

\subsection{ODE Model trained with Flow Matching}
\label{OMFM}
For a generative task, its ordinary differential modeling can be represented as follows \cite{lipman2022flow}:
    \begin{equation}
    \frac{d}{dt}\phi_t(x) = v_t(\phi_t(x));~~ \phi_0(x) = x
    \end{equation}
where $x$ denotes data points in data space $\mathbb{R}^d$, $t \sim U[0, 1]$, and $v_t$ is a learnable time-dependent vector field that allows us to flow reversely from the initial distribution $p_0$ to the target distribution $p_1$. A simple method to optimize the $v_t$ is as the following equation: 
    \begin{equation}
    \mathop{min}\limits_{\Theta} \int_{0}^{1} \mathbb{E} \left [||v_t(x) - u_t(x)||^2 \right ] dt\label{eq:2}
    \end{equation}
where $\Theta$ is the parameters of $v_t$, and $u_t$ is the vector field that generates forward trajectories $p_t(x)$. However, we have no explicit formulas for $p_t$ and $u_t$. 

Fortunately, Lipman et al. \cite{lipman2022flow} point out that if we take a non-causal perspective, given a random variable distributed $x_1$ from the target distribution, there is a non-causal conditional flow that corresponds to $u_t(x|x_1)$:
    \begin{equation}
        \varphi_t(x) = tx_1 + (1 - (1-\sigma_{min})t)x \label{eq:3}
    \end{equation}
where $\sigma_{min}$ approaches 0. This flow is also called the Optimal Transport (OT) displacement map between $p_0(x|x_1)$ and  $p_1(x|x_1)$. Then, the conditional vector field $u_t(x|x_1)$ takes the form:
    \begin{equation}
        u_t(x|x_1) = \frac{d}{dt}\varphi_t(x) = x_1 - (1-\sigma_{min})x \label{eq:4}
    \end{equation}
Reparameterizing it in terms of just $x_0$, which is a random variable distributed from the initial distribution, we can get a new optimization objective: 
    \begin{equation}
    \mathop{min}\limits_{\Theta} \int_{0}^{1} \mathbb{E} \left [||v_t(\varphi_t(x_0)) - (x_1-(1-\sigma_{min})x_0)||^2 \right ] dt\label{eq:5}
    \end{equation}
where the $(x_1-(1-\sigma_{min})x_0)$ is the direction of a straight line from $x_0$ to $x_1$. Previous work has demonstrated that a $v_t$ trained with Eq. (\ref{eq:5}) will give the causal direction of the reverse trajectories according to the marginal probability of forward trajectories. \cite{lipman2022flow,liu2022flow}. This method that trains neural ODE models with the OT trajectories is called flow matching. 

While the OT map moves in straight trajectories, the reverse trajectories predicted by $v_t$ are not always so because the numerical simulation of ODE is a causal process. Liu et al. \cite{liu2022flow, lee2023minimizing} demonstrated that the reverse trajectories of an ODE model trained with FM are usually curved. This curvature is caused by intersections in the OT map. We visualize this phenomenon with point transmission experiments in 2-dimensional space. As shown in Fig. \ref{fig1}, we trained two ODE models to generate red points from green points following two different distributions. When the initial points follow distribution B (Fig. \ref{fig1} (c)), since the $(x_0, x_1)$ pairs are randomly sampled during the training, a target point can map to any initial starting point, leading to significant crossovers in the forward trajectories. Such intersecting trajectories can limit the ODE model in two ways: 1) At the intersection of forward trajectories, an intermediate state $\varphi_t(x)$ can correspond to multiple completely different directions, which poses a greater learning challenge for the neural network. This directly leads to the network generating incorrect reverse trajectories in places where the forward trajectories do not exist, as depicted in the middle-upper portion of Fig. \ref{fig1} (d), which in turn generates undesirable samples. 2) Because of the uniqueness of the solution of the ODE, the predicted reverse trajectories only fit the shape of the forward trajectories in a non-intersecting form, resulting in the high curvature of reverse trajectories. Therefore, the ODE solver needs more NFEs to match the predicted trajectory without increasing the truncation error. In contrast, if the initial points follow distribution A (Fig. \ref{fig1} (a)), which can deterministically correspond to the final distribution, the intersections in the OT map will obviously reduce. In this case, the neural network can learn without perplexity, giving straighter reverse trajectories, as shown in Fig. \ref{fig1} (b). Since the direction of the reverse trajectory remains constant, the solver can match the entire reverse trajectory with extremely few NFEs. 

Extending from 2D point to speech, as far as we know, the previous works \cite{le2023voicebox,kim2024p,mehta2024matcha,guo2024voiceflow} based on ODE all assume the initial data $x_0 \sim \mathcal{N}(0, I)$. As we analyzed above, this default assumption could limit the performance of the ODE model trained with FM. Consequently, a more stable and reliable initial distribution is needed to enable ODE-based text-to-speech models to generate high-quality speech more effectively. The experimental results at Sec. \ref{ca} provide support for this claim.


\subsection{Voice Clone with modified ODE}
\label{OM}

Motivated by the requirement of initial feature that more deterministically coupled to acoustic features, we assume that speech contains three main components: 1) average pronunciation, 2) text-related prosody\footnote{Prosody determined only by the current and in-context text.}, 3) speaker information and speaker-related prosody\footnote{Prosody changes influenced by personal information such as timbre and accent.}. We only consider duration-independent prosody here since the phoneme duration is modeled by an additional model. 

Given the above assumptions, for the same sequence of average pronunciation $X = {\{}x_1,x_2,...,x_i{\}}$, there are a variety of possibilities for text-related prosody. Thus, an average pronunciation sequence could corresponds to different composite sequences:
    \begin{equation}
    Y~{\in}~{\{} X^{a_1},X^{a_2}, ..., X^{a_j} {\}}\label{eq:7}
    \end{equation} 
where $a_j$ represents the $j^{th}$ text-related prosody scheme. Similarly, the same sequence Y could be combined with different speaker information and speaker-related prosody. The combined sequence can be written as:
    \begin{equation}
    Z~{\in}~{\{} Y^{b_1},Y^{b_2}, ..., Y^{b_n} {\}}\label{eq:8}
    \end{equation} 
where $b_n$ represents the $n^{th}$ speaker information and related prosody. Although $Z$ does not contain all the speech information, it represents a unique speech audio more than random noise. In other words, $Z$ and speech are the deterministic coupling pair that can reduce the curvature of reverse trajectories learned by the ODE model.

To train the ODE model with this pair of data, SF-Speech generates speech through three steps, as shown in Fig. \ref{fig2}. First, we model the average pronunciation and text-related prosody jointly from phoneme sequences with a text encoder, as they are highly correlated with the text. The output of the text encoder, $F_1$, represents the sequence $Y$ in Eq. \ref{eq:7}. Then, a speaker adder is utilized to extract the speaker information and speaker-related prosody from the in-context prompt $F_p$ and add them to the $F_1$. The output of the speaker adder $F_2$ is considered as the embedding of the sequence $Z$ in Eq. \ref{eq:8}. Finally, an ODE module, called detail ODE, is employed  to transport $F_2$ to the mel-spectrograms. For convenience, we use $F_3$ later to represent the mel-spectrograms. To help the detail ODE converge stably, $F_1$, $F_2$, and $F_p$ are concatenated together as its input, as illustrated on the right of Fig. \ref{fig2}. Furthermore, to learn a continuous distribution from the limited training speech, we follow \cite{liu2022flow} and add a continuous noise $\xi$ on $F_2$ during training, where $\xi \sim \mathcal{N}(0,I)$. As a result, the optimization goal of SF-Speech can be written as:
\begin{equation}
    \label{eq:9}
    \begin{gathered}
    \mathop{min}\limits_{\Theta} \int_{0}^{1} \mathbb{E} \left [ \big|\big| v_t(\varphi_t(\widetilde{F_2}),F_1,F_2,F_p) - \frac{d}{dt}\varphi_t(\widetilde{F_2}) \big|\big|^2 \right ] dt
    \end{gathered} 
 \end{equation}
where $\varphi_t(\widetilde{F_2}) = tF_3+(1-(1-\sigma_{min})t)\widetilde{F_2}$ and $\widetilde{F_2} = F_2 + \xi$.

\subsection{Text Encoder}\label{af}

SF-Speech splits the training audio-text pairs into the predicted part and in-context prompt with a continuous mask matrix $m$. Given a phoneme sequence $y_{text} \in [V]^N$ of training text, where the $V$ means the number of phonemes appearing in the training set, a lookup table embedding it into $y_{emb} \in \mathbb{R}^{N\times H}$. Then, inspired by StyleTTS \cite{li2022styletts}, a 1D convolution neural network (CNN) is employed to fuse the local features of $y_{emb}$ and a bidirectional long-short term memory (Bi-LSTM) network follows to fuse its long-time information. We use a length regulator module \cite{ren2019fastspeech} upsamples this feature to $y^{'}_{emb} \in \mathbb{R}^{T\times H}$ with $T$ denoting the number of mel-spectorgrams frame, according to the phoneme duration. A regression duration model mentioned in \cite{le2023voicebox} is trained separately to predict the masked duration with the unmasked duration and $y_{text}$. Finally, only the masked part $(1-m)\odot y^{'}_{emb}$ participate as the text feature $F_1$ in the subsequent data stream.

\begin{figure*} [h]

        \begin{minipage}{0.43\linewidth}
            \centerline{\includegraphics[width=\textwidth]{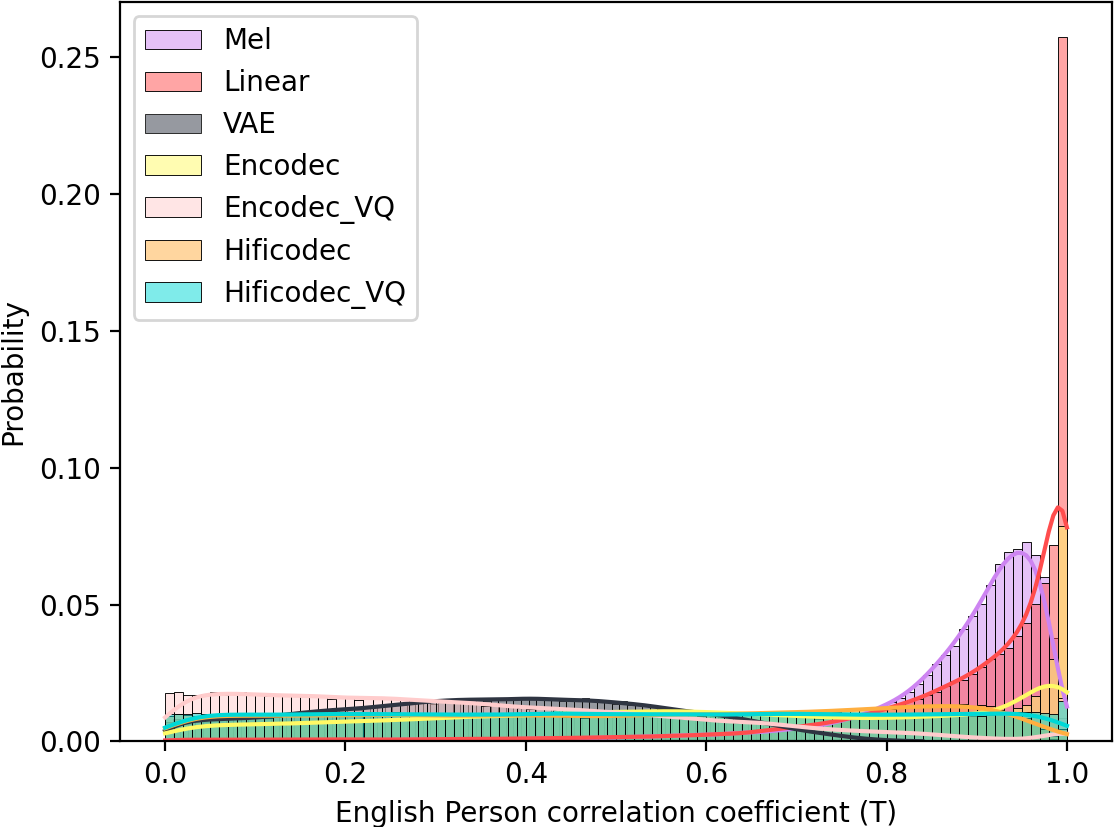}}
        \end{minipage}
        \begin{minipage}{0.43\linewidth}
            \centerline{\includegraphics[width=\textwidth]{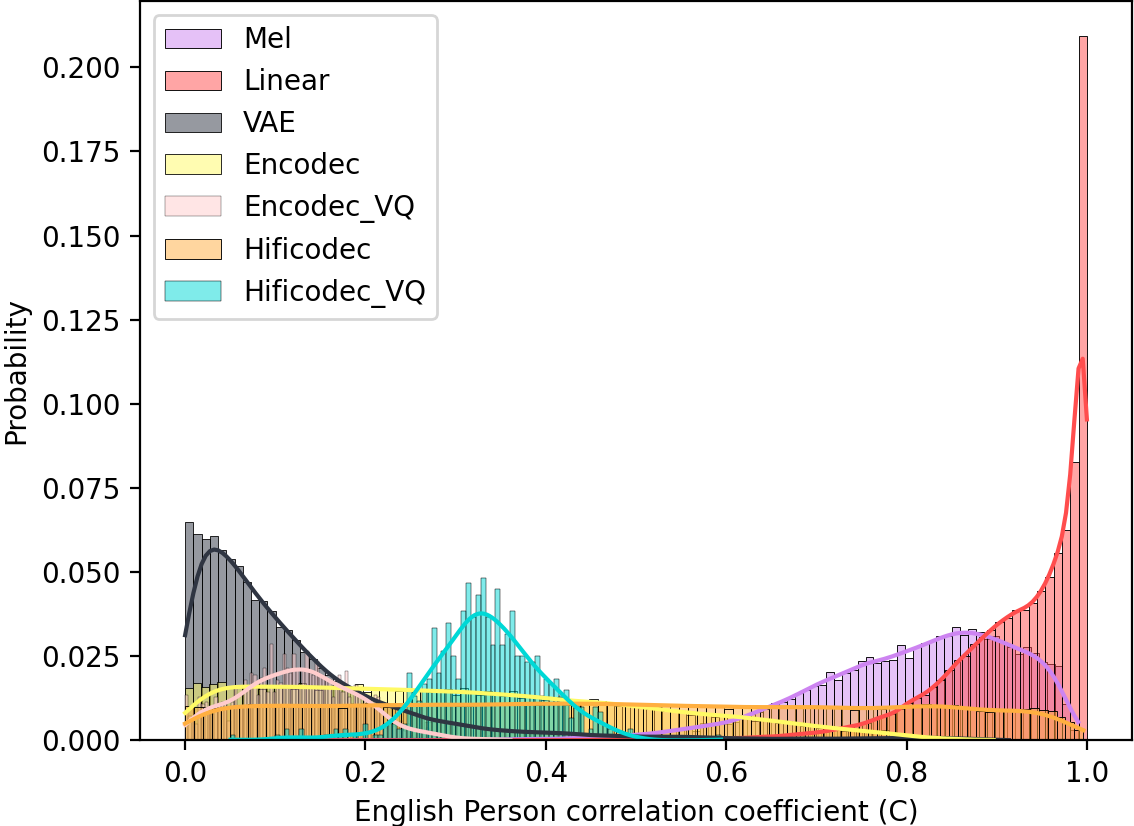}}
        \end{minipage}
        \begin{minipage}{0.43\linewidth}
            \centerline{\includegraphics[width=\textwidth]{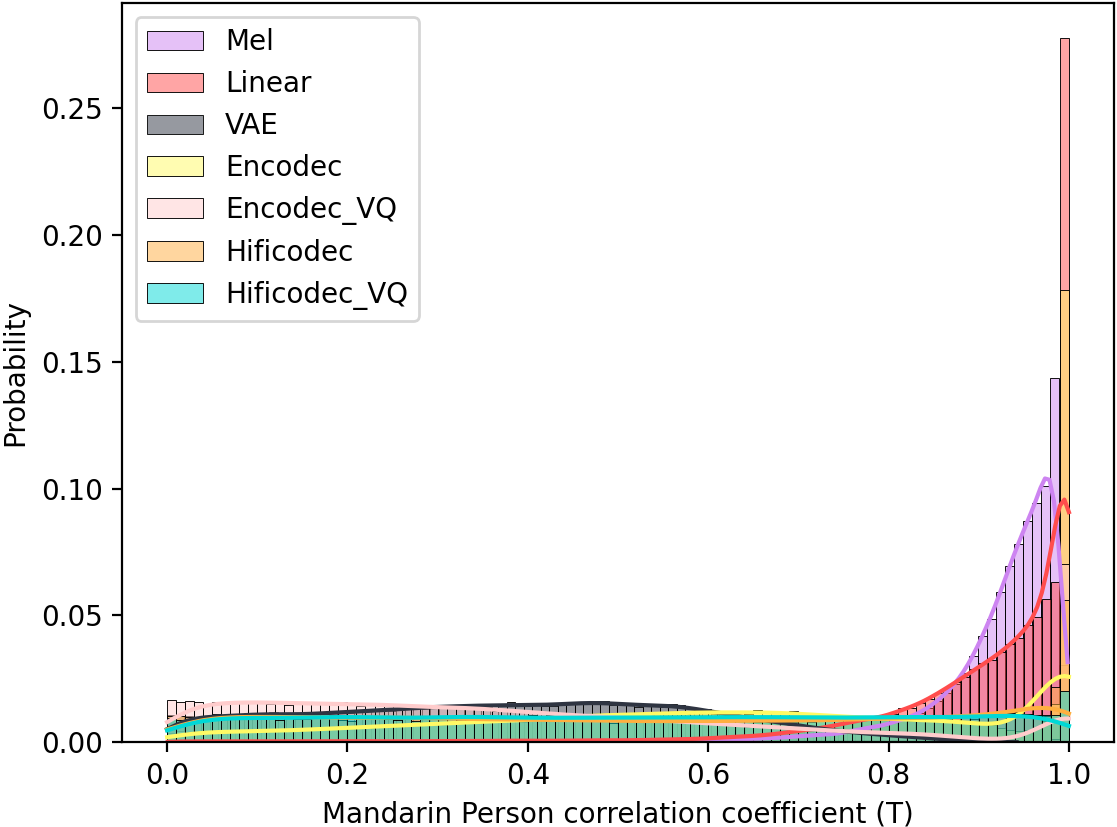}}
        \end{minipage}
        \begin{minipage}{0.43\linewidth}
            \centerline{\includegraphics[width=\textwidth]{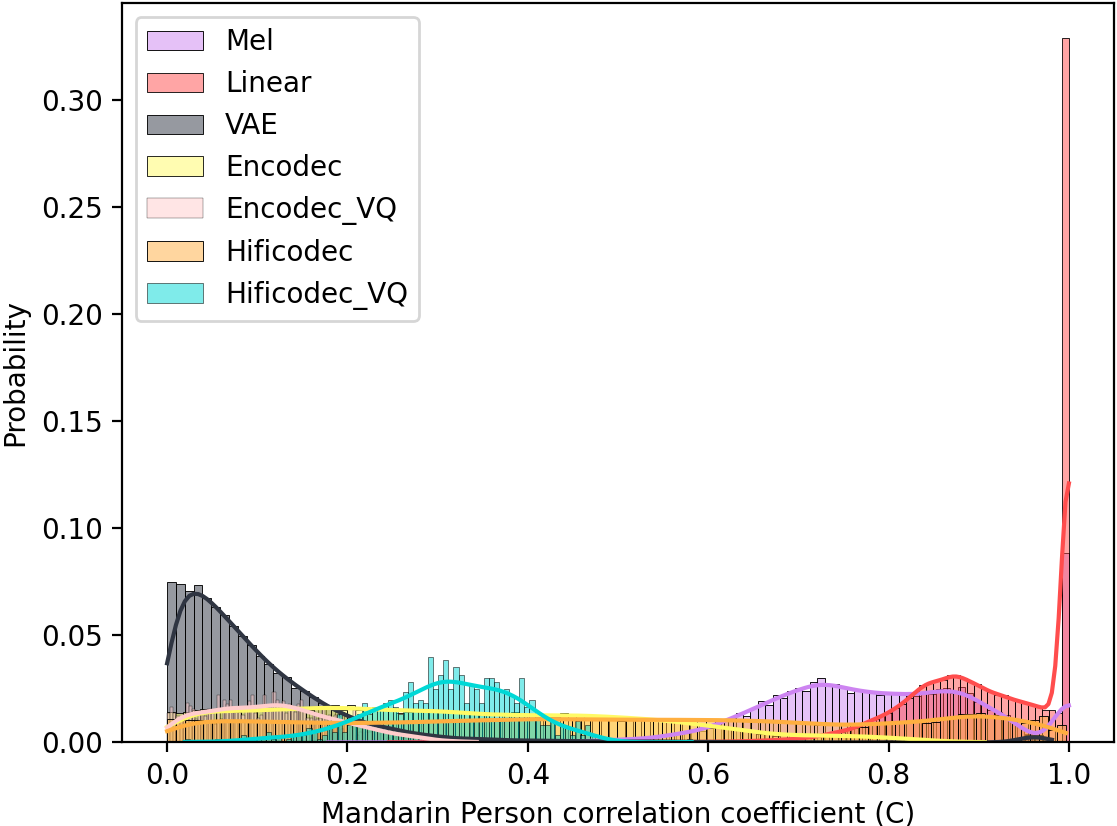}}
        \end{minipage}
\centering
\caption{The probability distribution of the PCCs-AV on time (T) and channel (C) axes for different features in English and Mandarin. "Mel" and "Linear" denote mel-spectrogram and linear spectrogram, respectively. "VAE" means the latent embedding from VITS \cite{kim2021conditional}. "Encodec" and "Hificodec" represent the latent embedding from EnCodec \cite{defossez2022high} and HiFi-codec \cite{yang2023hifi}. "\_VQ" stands for the indexes of the corresponding codebook.}
\label{fig3}
\end{figure*}

\subsection{Speaker Adder}\label{cf}
Given the remarkable effectiveness of in-context learning in zero-shot voice cloning, Speaker adder extracts speaker-related information from the in-context mel-spectrogram $F_p$. Specifically, we concatenate $F_p$ with $F_1$ on the hidden axis as the input of speaker adder, $z_{coa} \in \mathbb{R}^{T\times H}$. Like the text encoder, the speaker adder consists of CNN and the Bi-LSTM network. The difference is that this module utilizes the Bi-LSTM network before the CNN layer to encode the in-context speaker information into each frame. It is worth mentioning that because our goal is to find the distribution that deterministically coupled to mel-spectrograms in its space, the hidden dimension of the CNN network is progressively reduced until the $z_{coa}$ is transformed into $z^{'}_{coa} \in \mathbb{R}^{T\times C}$ with $C$ denoting the number of frequency channels of mel-spectrograms. We found that the output of the speaker adder in the trained SF-Speech looks like a coarse mel-spectrogram, even though no mel-spectrogram-related loss constraints are made for the text encoder and speaker adder, which is the reason why the $F_2$ is named as "coarse feature". Like the text feature, only the masked part $(1-m)\odot z^{'}_{coa}$ participates as the coarse feature $F_2$ for the subsequent processing.


\subsection{Detail ODE}
\label{sss}

In the field of diffusion generation, selecting a Transformer-based network as the backbone has become a common tendency \cite{le2023voicebox, peebles2023scalable, bao2023all}. However, in our experiments, we observed that employing the network lacking the convolution layer as the direction estimator of ODE-based models leads to generated mel-spectrograms lacking coherence, which also was observed in U-Dit TTS \cite{jing2023u}, a SDE-based model. Interestingly, we also found that apply the same network to neural-coded latent acoustic features instead of mel-spectrograms resulted in the generated speech sounding normal. Hence, we speculate that there is a difference in local correlation between mel-spectrograms and those neural-coded latent features, leading to convolution being more friendly to modeling the former. To validate this presumption, we visualized the probability distribution of the Pearson correlation coefficients between adjacent vectors (PCCs-AV) at time and channel axes for different acoustic features. Fig. \ref{fig3} shows the results for 7 acoustic features on 400 samples across English and Mandarin, which are from 6 datasets, including  VCTK \cite{2017CSTR}, LibriSpeech \cite{panayotov2015librispeech}, LJSpeech \cite{ljspeech17}, Expresso \cite{nguyen2023expresso}, CSMSC \cite{csmsc17}, and Aishell-3 \cite{shi2020aishell}. These results illustrate that the mel and linear spectrograms exhibit significantly stronger local correlation in both the time and channel domain compared to those latent features extracted by neural models. This phenomenon explains the discovery we mentioned above. Moreover, it offers statistical backing to prior models \cite{gong2022ssast,guo2022multi,ren2019fastspeech, gulati2020conformer}, that improved their performance using 1D or 2D CNN in mel-spectrogram-related tasks but did not analyze the reasons in depth.  

\begin{figure}
\centering
\begin{minipage}[b]{.99\linewidth}
\includegraphics[width=.99\linewidth]{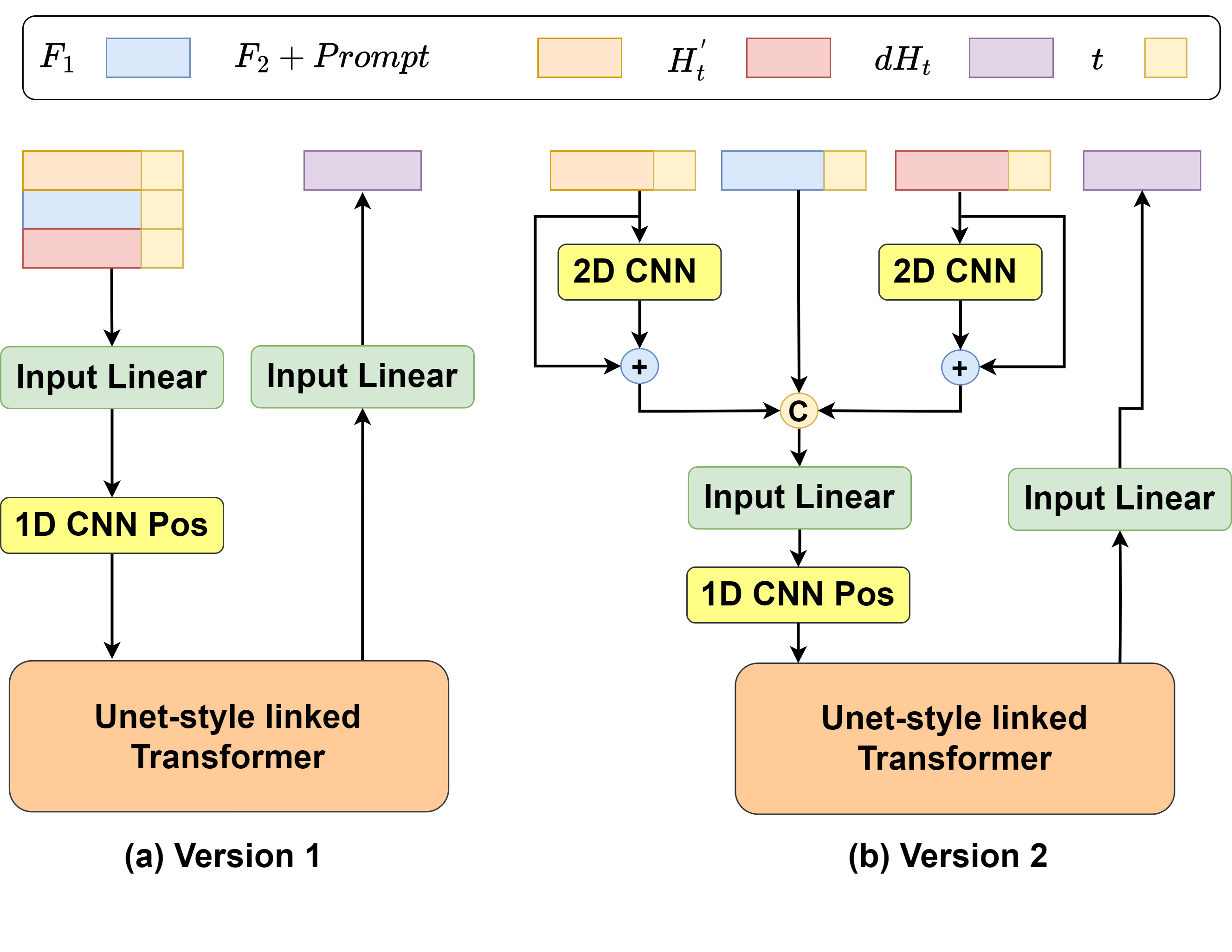}
\end{minipage}
\caption{Two versions of detail ODE structure. Version 1 consists of the Unet-style linked Transformer and 1D convolutional positional embedding. Version 2 combines 2D convolutional layers based on Version 1.}
\label{fig4}
\end{figure}

Inspired by this, we considered two versions of network structures for the detail ODE, as illustrated in Fig \ref{fig4}. The first version follows the structure in VoiceBox \cite{le2023voicebox} containing the Transformer with Unet-style link \cite{bao2023all} and a 1D convolutional positional embedding layer. We use this version to compare SF-Speech with ODE-based baseline models as fairly as possible. The second version uses two 2D depthwise separable convolution layers to fuse local features of $F_2 + F_p$ and $\varphi_t(\widetilde{F_2})$ before entering them into the Transformer. The text feature $F_1$ does not go through the 2D convolutional layer because its channel axis contains no frequency information. The comparison results of these two network structures can be found in Sec. \ref{nc}.



\section{Experiments}
\label{ee}
In this section, we set up a series of experiments to answer three main questions: 1) Can SF-Speech outperform other large-scale speech generation models in the voice cloning task? 2) Can SF-Speech maintain its relative superiority over other models when the data scale drops significantly? 3) Does the design of the initial distribution in SF-Speech straighten the reverse trajectory, and how does this straightening flow benefit TTS? The experimental configuration related to datasets, implementation details, and baselines is presented as follows.



\subsection{Dataset}
\label{dataset}

\textbf{Large-scale dataset:} To answer the first question, we conduct partial experiments on Emilia \cite{he2024emilia}. Emilia is constructed on in-the-wild data using an automated processing pipeline, encompassing over 101K hours of data across six languages. We selected the English and Chinese data from this dataset and further filtered it following the approach in F5-TTS \cite{chen2024f5}. The resulting high-quality subset contains approximately 100K hours of speech.

\textbf{Small-scale dataset:} To answer the second question, we perform small-scale experiments on MagicData \cite{magic19}, a Chinese speech dataset recorded by a diverse range of non-specialized devices. This dataset, which has not undergone the enhancement pipeline, closely mimics the characteristics of real-world speech. Comprising a total of 755 hours of speech, it was recorded by 1000 speakers from different accent regions across China. This unique composition of MagicData enables us to effectively investigate the performance of our model under small-scale and in-the-wild data conditions.

\textbf{Test data:} We randomly selected 32 unseen speakers (16 females, 16 males) from the MagicData, with 15 utterances for each speaker as the Chinese zero-shot text-to-speech (ZS-TTS) test data. We also build a speech restoration (SR) test set containing 4000 samples from MagicData. For the English ZS-TTS test, we use the LibriSpeech-PC test-clean released by F5-TTS \cite{chen2024f5}.

We transcript the text to phonemes with Phonemizer \cite{Bernard2021} and align them with speech using MFA \cite{2017Montreal}. A pre-trained BigVGAN\footnote{https://huggingface.co/nvidia/bigvgan\_base\_22khz\_80band} \cite{lee2022bigvgan} is used for experiments on MagicData and a pre-trained Vocos\footnote{https://huggingface.co/charactr/vocos-mel-24khz} \cite{siuzdak2023vocos} is employed for experiments on Emilia.

\subsection{Implementation Details}
\label{ID}

We set two parameter scales for SF-Speech to correspond to various-scale datasets. Both of them have the same text encoder and speaker adder. The CNN of the text encoder consists of three 1D convolutional layers with layer normalization, and its LSTM network consists of one Bi-LSTM layer. The hidden dimensions of Bi-LSTM and CNN are 512, and CNN's kernel size is 5. The LSTM network of the speaker adder also consists of one Bi-LSTM layer with 256 hidden dimensions, while its CNN consists of seven 1D convolutional layers with instance normalization \cite{ulyanov2016instance}. The kernel size of CNN in the speaker adder is set to 3 and the hidden dimension of the top four layers is set to 512, the middle two layers to 256, and the last to 80. The Unet-style linked Transformer of the detail ODE has 16 attention heads and the embedding/FFN dimension of 1024/4096, following the VoiceBox \cite{le2023voicebox}. We employ 8 transformer layers for the small SF-Speech and 24 layers for the large one. The 2D CNN in version 2 of detail ODE has an embedding dimension of 256 and a kernel size of $3\times3$. Unless otherwise written, SF-Speech uses version 1 by default to compare with other ODE-based models under the same network structure. 

We apply the masked training and classifier-free guidance \cite{ho2022classifier} in \cite{le2023voicebox} to the proposed model. Only the masked frames are considered when computing the loss. An AdamW optimizer with a peak learning rate of 2e-5, linearly warmed up for 5000 steps and decayed in cosine annealing over the rest steps, was employed to train the small SF-Speech for 500K iterations on 2 NVIDIA V100 GPUs. The large SF-Speech was trained to 1.2M steps with the same optimizer as mentioned in \cite{chen2024f5} but a smaller batch size of 76800 audio frames (0.23 hours) on 4 NVIDIA A100 GPUs.

\subsection{Baselines}

Seven baseline models are compared in our experiments to evaluate the proposed method on zero-shot voice clone tasks. The details of them are as follows:

\begin{itemize}
\item \textbf{YourTTS:} YourTTS \cite{casanova2022yourtts} is a classic GSE-based model in zero-shot voice clone tasks. It employed VITS \cite{kim2021conditional} as its backbone and introduced the speaker embedding extracted by H/ASP \cite{heo2020clova}. We cloned the corrected official implementation\footnote{https://github.com/coqui-ai/tts} and trained it for 1140K steps with the default configuration.

\item \textbf{VoiceBox} and \textbf{VoiceBox-S:} VoiceBox \cite{le2023voicebox} is the SOTA ODE-based zero-shot model. We implemented it and set up two different scales. The first scale follows the 24-layer Transformer in the original paper with 330M parameters. The second scale uses the 8-layer Transformer mentioned in Sec. \ref{ID} with 110M parameters called VoiceBox-S. VoiccBox and VoiceBox-S were trained for 500k steps using the same training strategy as SF-Speech. The same BigVGAN was used as their companion vocoder to ensure a fair comparison.

\item \textbf{VALL-E:} VALL-E is the first autoregressive zero-shot voice clone model based on in-context learning and LLM. We cloned an unofficial implementation\footnote{https://github.com/lifeiteng/vall-e} of it and trained it for 600K steps with prefix mode 1 until the valid loss stopped dropping.

\item \textbf{E2 TTS:} E2 TTS \cite{eskimez2024e2} is an improved version of Voicebox. It adopts the Transformer-based network in Voicebox but abandons the explicit modeling of phoneme durations. We employ the checkpoint\footnote{https://huggingface.co/SWivid/E2-TTS/tree/main} trained on Emilia in our experiments.

\item \textbf{SimpleSpeech2:} SimpleSpeech2 \cite{yang2024simplespeech} is a concurrent work of our study that applies the ODE model to the latent space and also uses the sentence duration to control speech generation. It achieved performance comparable to that of large-scale TTS models on a small-scale dataset. In order to conduct an accurate comparison, we reached out to the authors and obtained the test samples.

\item \textbf{F5-TTS:} F5-TTS \cite{chen2024f5} is another concurrent work that modifies the network structure in E2 TTS, allowing faster inference. We employ the official checkpoint\footnote{https://huggingface.co/SWivid/F5-TTS/tree/main/F5TTS\_Base} trained on Emilia in our experiments.

\end{itemize}

\subsection{Evaluation Metrics}
We evaluate our model with two aspects: speech quality and timbre similarity. The subjective and objective evaluation metrics for each aspect are as follows:

\textit{1) Speech Quality:}
\begin{itemize}
\item \textbf{QMOS} We set a Quality Mean Opinion Score (QMOS) test to evaluate the speech quality. Fifteen native speakers were asked to score between 1 and 5 as the Absolute Category Rating(1: Bad, 2:Poor, 3:Fair, 4:Good, 5:Excellent), and the minimum score interval allowed was 0.5. The final result is presented in the form of a 95\% confidence interval.
\item \textbf{WER} We use the Word Error Rate (WER) of the Automatic Speech Recognition (ASR) system to measure the intelligibility of generated speech as previous works \cite{le2023voicebox, wang2023neural}. Given that the speech used for experimentation is Chinese, an open-source WeNet \cite{zhang2022wenet} with the pre-trained checkpoint "multi\_cn"\footnote{https://github.com/wenet-e2e/wenet/tree/main/examples/multi\_cn/s0} is used as our ASR system. 
\item \textbf{DNSMOS} Considering that the quality of the generated speech needs to be repeatedly evaluated many times in Sec. \ref{iee}, we use DNSMOS P. 835 \cite{reddy2022dnsmos} instead of the subjective QMOS to evaluate the speech generated with the different number of function evaluations (NFEs) automatically.  
\end{itemize}

\textit{2) Timbre Similarity:}
\begin{itemize}
\item \textbf{SMOS} We set a Similarity Mean Opinion Score (SMOS) test to score the timbre similarity between the prompt and generated speech. The scoring rules are the same as those for QMOS.
\item \textbf{SIM} We employ ECAPA-TDNN \cite{desplanques2020ecapa}, a speaker recognition model, to extract speaker embedding of reference and generated speech. The cosine similarity between two embeddings is used to evaluate the timbre similarity objectively. SIM-o represents the similarity to the original reference audio, and SIM-r reports the similarity to the reconstructed reference audio by the vocoder.
\end{itemize}

\section{Experiment results}
\label{er}

In this part, we first discuss the experimental results on large-scale datasets in Sec. \ref{lsdr}. Subsequently, in Sec. \ref{ssdr}, we present the results on small-scale datasets. The inference efficiencies of different models are then evaluated in Sec. \ref{iee}. Sec. \ref{ca} conducts a deeper analysis of the flow curvature in ODE-based models and shows the advantages of straightened flow. Finally, we report the ablation results of the network structure and features in SF-Speech to illustrate the impact of each component.


\begin{table*}[htbp]
  \begin{center}
  \caption{Subjective and objective evaluation results in Chinese test. $\heartsuit$ denotes the model uses the checkpoint released by the official repository of F5-TTS. $\clubsuit$ means the inference samples from authors. $\diamondsuit$ means our reproduction.}
  \label{tab1}
  \begin{tabular}{c|c|c|c|c c c c c} 
  \toprule [2pt]
  \textbf{Task\&Test set} & \textbf{Model} &\textbf{Data(hrs)} & \textbf{Parameters} & \textbf{QMOS} & \textbf{SMOS} & \textbf{WER} & \textbf{SIM-o} & \textbf{SIM-r} \\
  \midrule
  \multicolumn{4}{c|}{Ground truth}    & 4.07 $_{\pm 0.05}$ & n/a & 6.22 & n/a & n/a \\
  \midrule
  \multirow{10}{*}{\makecell{ZS-TTS \\ \& \\ MagicData test}} 
            & E2 TTS(32 NFEs)$\heartsuit$  &  100K EN+CH  &333M &3.67 $_{\pm 0.06}$ & 4.05 $_{\pm 0.05}$ & 15.02& 0.646 & 0.651\\
            & F5-TTS(32 NFEs)$\heartsuit$  &  100K EN+CH  &336M &3.98 $_{\pm 0.05}$ & 3.97 $_{\pm 0.07}$ & 8.34 & 0.609 & 0.617\\
            & VoiceBox(32 NFEs)$\diamondsuit$  & 100K EN+CH  &330M &3.91 $_{\pm 0.06}$ & 3.93 $_{\pm 0.09}$ & 8.41 & 0.633 & 0.640 \\
            & SF-Speech(8 NFEs)  & 100K EN+CH & 337M &\textbf{4.06 $_{\pm 0.08}$} & \textbf{4.12 $_{\pm 0.03}$} & \textbf{4.84} & \textbf{0.665} & \textbf{0.671}\\
    \cmidrule{2-9}
            & SimpleSpeech2(25 NFEs)$\clubsuit$  & 11K EN+CH & 345M &3.61 $_{\pm 0.03}$ & 3.54 $_{\pm 0.04}$ & 20.04 & 0.502 & -\\
    \cmidrule{2-9}
            & YourTTS$\diamondsuit$  &  0.7K CH &87M &3.09 $_{\pm 0.08}$ &3.52 $_{\pm 0.06}$&31.66 & 0.509 & n/a\\
            & VALL-E$\diamondsuit$   &  0.7K CH& 367M    &3.04 $_{\pm 0.10}$      &3.15 $_{\pm 0.09}$&  27.04     & 0.287 &0.336 \\
            & VoiceBox-S(8 NFEs)$\diamondsuit$ &  0.7K CH& 110M  & 3.52 $_{\pm 0.07}$&3.61 $_{\pm 0.07}$&15.99  & 0.516 &0.524\\
            & VoiceBox(8 NFEs)$\diamondsuit$   &  0.7K CH& 330M & 3.66 $_{\pm 0.05}$&3.88 $_{\pm 0.07}$&12.62 & 0.543 & \textbf{0.554} \\
            & SF-Speech-S(8 NFEs)  &  0.7K CH& 117M &\textbf{3.71 $_{\pm 0.04}$}& \textbf{3.91 $_{\pm 0.05}$}&  \textbf{12.41} & \textbf{0.545} &\textbf{0.554}\\
  \midrule
  \multirow{3}{*}{\makecell{SR \\ \& \\ MagicData test}} 
  & VoiceBox-S(8 NFEs)$\diamondsuit$ &  0.7K CH& 110M  &3.45 $_{ \pm 0.06}$&3.71 $_{ \pm 0.10}$ & 10.76 &0.681 & 0.714\\
  & VoiceBox(8 NFEs)$\diamondsuit$  &  0.7K CH & 330M &3.69 $_{ \pm 0.06}$&\textbf{4.00 $_{ \pm 0.04}$ }& 10.68 &\textbf{0.696} & \textbf{0.728}\\
  & SF-Speech-S(8 NFEs)  &  0.7K CH & 117M   & \textbf{3.73 $_{ \pm 0.05}$}  &3.97 $_{ \pm 0.07}$& \textbf{8.69} &0.695 & 0.726\\
  \bottomrule [2pt]
  \end{tabular}
  \end{center}
\end{table*}

\begin{table*}[htbp]
  \begin{center}
  \caption{Subjective and objective evaluation results in English test. $\heartsuit$ denotes the model uses the checkpoint released by the official repository of F5-TTS. $\diamondsuit$ means our reproduction. $\dagger$ denotes the result reported in baseline papers.}
  \label{tab2}
  \begin{tabular}{c|c|c|c|c c c c c} 
  \toprule [2pt]
  \textbf{Test set} & \textbf{Model} &\textbf{Data(hrs)} & \textbf{Parameters} & \textbf{QMOS} & \textbf{SMOS} & \textbf{WER} & \textbf{SIM-o} & \textbf{SIM-r} \\
  \midrule
  \multicolumn{4}{c|}{Ground truth}  & 4.15 $_{\pm 0.04}$ & n/a & 2.23$\dagger$ & n/a & n/a \\
  \midrule
  \multirow{3}{*}{\makecell{LibriSpeech \\ test-clean}} 
            & YourTTS & 60K EN&  87M & - & - &7.7$\dagger$ & 0.337$\dagger$ & n/a\\
            & VALL-E & 60K EN & -    & - & - &5.9$\dagger$  & -     &0.580$\dagger$ \\
            & VoiceBox(32 NFEs) & 60K EN & 330M & - & - &1.9$\dagger$  & 0.662$\dagger$  &0.681$\dagger$\\
  \midrule
  \multirow{5}{*}{\makecell{LibriSpeech-PC \\ test-clean}} 
            & SimpleSpeech2(25 NFEs)$\clubsuit$  & 11K EN+CH & 345M &3.93 $_{\pm 0.04}$ & 3.53 $_{\pm 0.08}$ & 4.98 & 0.583 & -\\
            \cmidrule{2-9}
            & E2 TTS(32 NFEs)$\heartsuit$ & 100K EN+CH & 333M & 3.77 $_{\pm 0.03}$ & 3.96 $_{\pm 0.04}$ &2.95$\dagger$  & 0.69$\dagger$  &0.702\\
            & F5-TTS(32 NFEs)$\heartsuit$ & 100K EN+CH & 336M & \textbf{4.00 $_{\pm 0.05}$} & 3.80 $_{\pm 0.06}$ &2.42$\dagger$  & 0.66$\dagger$  &0.675\\
            & VoiceBox(32 NFEs)$\diamondsuit$  & 100K EN+CH & 330M & 3.82 $_{\pm 0.04}$ & 3.85 $_{\pm 0.07}$ & 3.20 & 0.671 & 0.688\\
            &SF-Speech(8 NFEs)  & 100K EN+CH & 330M & 3.98 $_{\pm 0.04}$ & \textbf{4.06 $_{\pm 0.05}$} & \textbf{2.34} &\textbf{0.709} & \textbf{0.723}\\
            
  \bottomrule [2pt]
  \end{tabular}
  \end{center}
\end{table*}

\subsection{Large-scale Dataset Results}
\label{lsdr}

The upper part of Tab. \ref{tab1} and Tab. \ref{tab2} show the ZS-TTS results of those models trained with large-scale datasets on the MagicData and LibriSpeech-PC test sets. SF-Speech achieved the highest speaker similarity scores in both Chinese and English tests. Compared with E2 TTS, which employs the same network structure as the detail ODE in SF-Speech but ranks second in speaker similarity evaluation, SF-Speech obtains improvements of 0.15(CH) and 0.19(EN) in the SIM-o, and enhancements of 0.19(CH) and 0.10(EN) in the SMOS, with only 8 NFEs. These results indicate that the initial distribution in SF-Speech significantly reduces the modeling complexity and straightens the reverse trajectory. Turning the perspective to speech quality and intelligibility, SF-Speech gains a WER of 4.84\% and a QMOS of 4.06 in the Chinese test, representing a 42.0\% relative decrease and a 2\% relative increase compared to F5-TTS, the second-best model. However, in the LibriSpeech-PC clean test, the edge of SF-Speech over F5-TTS, E2 TTS, and SimpleSpeech2 diminishes, with the best WER of 2.34\% and a comparable QMOS of 3.98. We believe this is because the Chinese test data is closer to in-the-wild data, while the English test data predominantly features formal, read-aloud speech styles. We observed the methods with implicit alignment modeling exhibit inherent deficiency robustness when facing spontaneous speech in subjective evaluation. It is also worth noting that, although SF-Speech demonstrated better performance in the experiments on Emilia, it could not have as favorable training as F5-TTS and E2 TTS, with a smaller batch size and fewer GPUs.


\subsection{Small-scale Dataset Results}
\label{ssdr}
The results on small-scale datasets are shown at the bottom of Tab. \ref{tab1}.  Compared with the results on large-scale datasets, we observe various performance degradations across different models. We find that YourTTS and VALL-E perform significantly worse than their versions trained with large-scale datasets. For VALL-E, since the correlation between discrete acoustic feature and phoneme sequences drops a lot, 700 hours of training data is not enough to help it establish a stable mapping between these two sequences. We notice the spectrogram of speech generated by VALL-E contains artifacts in its high-frequency part, which have little effect on human hearing but are never seen by the speaker recognition model. This explains why VALL-E differs significantly from other models in the SIM test. In contrast, 700 hours of training data is sufficient for YourTTS which is based on VITS. However, we note that the prosody of speech generated by YourTTS significantly differs from real speech, which explains why it performs the worst on WER. We believe this can be attributed to the fact that YourTTS relies only on a global embedding to learn how to model the complex prosody of a speaker with content.

In addition, ODE-based models trained with MagicData also perform differently from those trained with Emilia. We notice that as the NFEs increase, WER and SIM are not continuously optimized. This is why we uniformly used NFEs of 8 to solve the ODE model trained using MagicData. The detailed analysis of this phenomenon is provided in Sec. \ref{iee}. Furthermore, when compared with YourTTS, VALL-E, and Voicebox-S trained with the same small-scale data, SF-Speech-S maintains a significant performance edge, with a WER of 12.41\% and a SIM-o of 0.545. Even when the parameter number of VoiceBox is increased to 330M, our method still stays on par with it across all metrics. Moreover, SF-Speech-S also outperforms SimpleSpeech2, which is trained with a larger dataset. These results demonstrate the effectiveness of SF-Speech on small-scale data. We also assessed VoicBox and SF-Speech-S on SR tasks, where we kept 15\% of the frames at both ends of the audio and masked the middle 70\% to reconstruct. In this evaluation, SF-Speech-S achieves the best QMOS(3.73) and WER(8.69\%) but competitive similarity scores. Notably, VoiceBox does not bring a considerable improvement on WER after tripling the number of parameters, while the proposed method breaks this bottleneck with only a 7M parameter increase. This suggests the predicament in model training engendered by the random noise initialization, as we analyzed in Sec. \ref{OMFM}.


%



\begin{table}[htbp]
  \begin{center}
  \caption{The RTF and model size of different models on the ZS-TTS task. $\heartsuit$ denotes the model uses the checkpoint released by the official repository of F5-TTS. $\diamondsuit$ means our reproduction. RTF1 is the result reported in baseline papers. RTF2 represents the result computed on our device.}
  \label{tab3}
  \begin{tabular}{c|c c c} 
  \toprule [2pt]
\textbf{Model} & \textbf{Model-Size(M)} & \textbf{RTF1} & \textbf{RTF2} \\
  \midrule
            VALL-E$\diamondsuit$         &367&0.62 &0.35 \\
            E2 TTS(32NFEs)$\heartsuit$ &333&0.68 &0.33 \\
            F5-TTS(32NFEs)$\heartsuit$ &336& 0.31&0.23\\
            VoiceBox(32NFEs)$\diamondsuit$   &330&0.64 &0.34\\
            SF-Speech(32NFEs)  &337& - &0.37\\
            SF-Speech(8NFEs)    &337& - &0.09\\
  \bottomrule [2pt]
  \end{tabular}
  \end{center}
\end{table}

\begin{figure*} [h]

         \begin{minipage}{1.\linewidth}
            \centerline{\includegraphics[width=\textwidth]{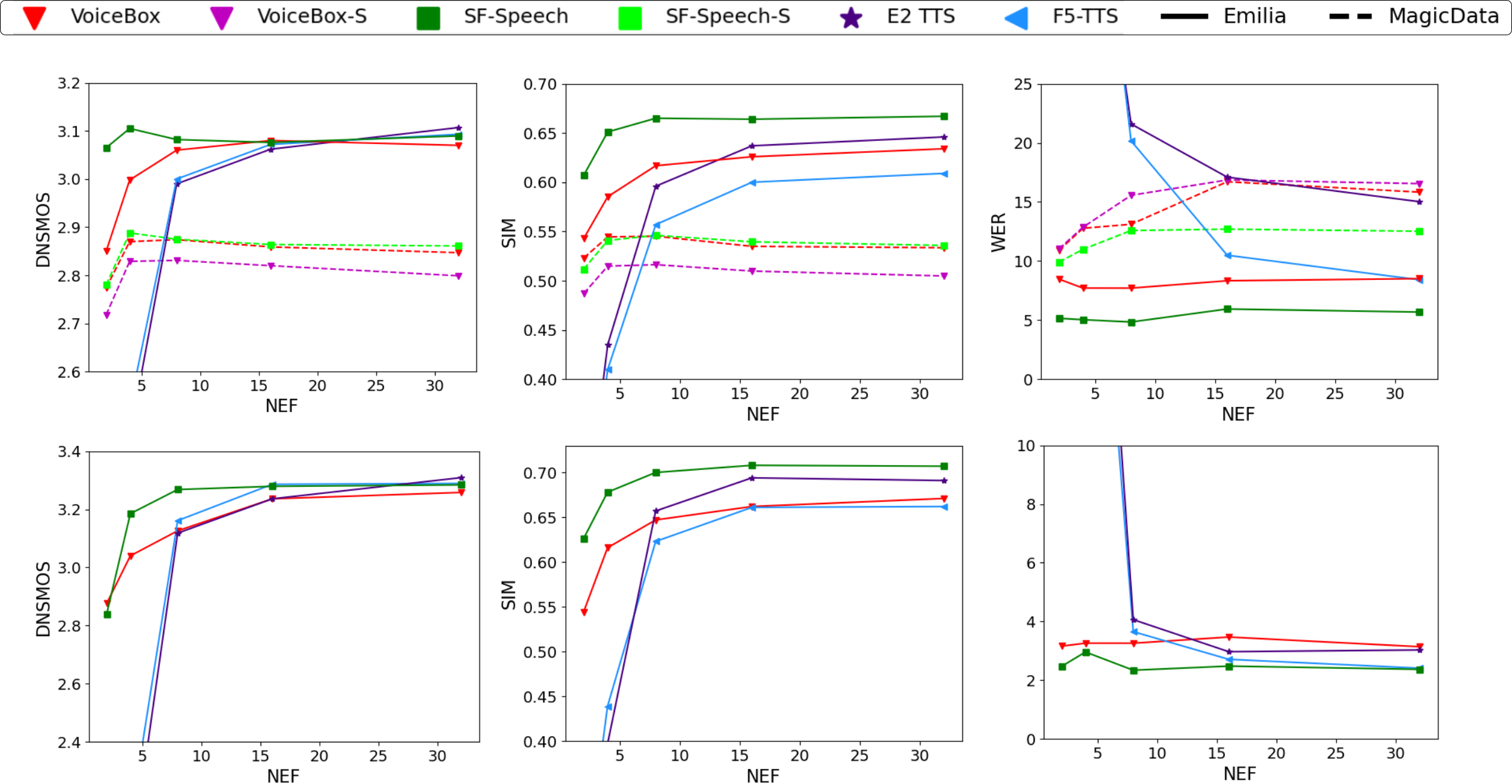}}
        \end{minipage}
        
\centering
\caption{Objective metrics of generated speech at different NFEs in ZS-TTS test. The first row shows the results on the MagicData test set, and the second row presents results on the LibriSpeech-PC test-clean set. The solid line means models trained on Emilia, while the dotted line represents models trained on MagicData.}
\label{fig5}
\end{figure*}

\subsection{Inference Efficiency Evaluation}
\label{iee}
In this part, we first measure the Real-Time Factor (RTF) for different models with a 10s inference speech and a Nvidia-A100 GPU. Then, we measure objective metrics under different NFEs for ODE-based models and give the optimal NFEs. 

\textit{1) RTF results:} 

Tab. \ref{tab3} displays the results measured on our device (RTF2) and those reported in previous papers (RTF1), where the trends between them are essentially consistent. When the NFEs of the ODE solver are set to 32, E2TTS, Voicebox, and SF-Speech exhibit close inference speeds. This is because their network architectures are nearly identical. The proposed model achieves an RTF2 of 0.37, which is merely 0.03 higher than that of VoiceBox, despite having two additional modules incorporating Bi-LSTM. F5-TTS exhibits the fastest inference speed, registering an RTF of 0.23. This is attributed to the substitution of certain transformer layers with ConvNeXT V2 blocks in its backbone. Furthermore, when the NFEs is reduced to 8, SF-Speech attains an RTF of 0.09, which is less than half that of F5-TTS (32 NFEs). Notably, in this case, SF-Speech can still generate the desired audio, while F5-TTS can not.


\begin{figure} [h]
    \begin{minipage}{0.66\linewidth}
        \centerline{\includegraphics[width=\textwidth]{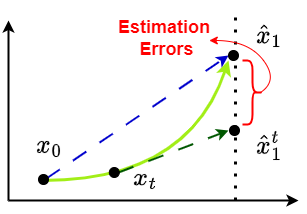}}
    \end{minipage}
\centering
\caption{The optimal transport direction (blue) and the real reverse trajectory (green).} 
\label{fig6}
\end{figure}

\textit{2) NFEs results:} 

We measure the objective metrics for all ODE-based models with the NFEs set at 2, 4, 8, 16, and 32. As shown by the solid lines in Fig.\ref{fig5}, these ODE-based models trained with Emilia exhibit different performance degradations as the NFEs decrease. F5TTS and E2TTS experience a substantial decline across all the metrics when NFEs are less than 8, while VoiceBox and SF-Speech are still able to generate intelligible audio in these cases. Moreover, we observe that SF-Speech always approaches the final scores(NFE=32) earlier than other models in DNSMOS and SIM tests. These results align precisely with the curvature of their reverse trajectory measured in Sec. \ref{ca}. In addition, there is no significant trend in the WER of SF-Speech and VocieBox, which is consistent with the results in \cite{le2023voicebox}. 

On the other hand, the VoicBox and SF-Speech trained with MagicData exhibit different trends from those trained with Emilia. As the NFEs increases, their SIM and DNSMOS decline slightly, and WER increases. We think this is because as the reverse trajectories are matched more accurately, these models add more audio imperfections (environmental noise, human noise, distortion due to recording devices, etc.) to the generated mel-spectrogram. These imperfections are widely present in the in-the-wild data of MagicData but are rare in Emilia. This is why we use the solver with NFEs of 8 to solve the ODE-based models trained with MagicData. Furthermore, our model still shows more stable metrics when NFE is greater than 8, suggesting a straighter generation trajectory.



\subsection{Curvature Analysis}
\label{ca}

\begin{figure} [h]
    \begin{minipage}{0.98\linewidth}
        \centerline{\includegraphics[width=\textwidth]{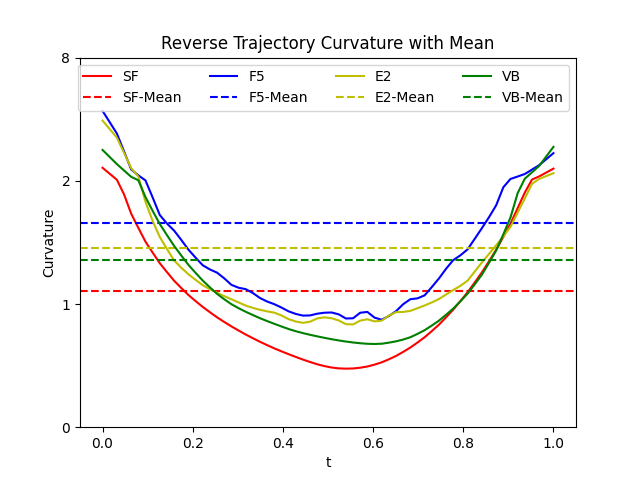}}
    \end{minipage}
\centering
\caption{The reverse trajectory curvature of different ODE-based models trained with Emilia.} 
\label{fig7}
\end{figure}

We quantify the curvature of the reverse trajectory by calculating the relative error between the predicted direction $v_t(x_0)$ and the optimal transport direction. As depicted in Fig. \ref{fig6}, the optimal transport direction is expressed as $\hat{x}_1 - x_0$, where the $\hat{x}_1$ is the terminal of the reverse trajectory simulated with a sufficiently large NFEs (128). The $\hat{x}_1$ is expected to be an ideal mel spectrogram that incorporates the timbre of an unseen speaker and the given content when the model is trained sufficiently. We measured 2000 reverse generation trajectories for those ODE-based models trained with Emilia. The results at different reversal time step $t$ and their average value are presented in Fig. \ref{fig7}. It can be seen that due to using distributions with deterministic coupling for FM training, the curvature of SF-Speech is significantly lower than that of other models, particularly during the early stages of the generation. This implies that with limited NFEs, the numerical solution of SF-Speech can be closer to $\hat{x}_1$. 

Moreover, when we relate this result to Fig. \ref{fig5}, which shows the objective metrics of these models under different NFEs, we find that the curvature at the early time step in the generation process is more crucial for inference efficiency. Compared with SF-Speech, those models characterized by larger curvature at the early time step show more significant declines across all metrics as the NFEs decreases. This is due to the fact that the large curvature at the early time step makes it impossible for the ODE solver with a small NFEs to precisely match the early trajectory, generating an intermediate state $x_t$ that deviates from the actual reverse trajectory. This incorrect $x_t$ then exacerbates the cumulative error in the subsequent generation process, leading to undesirable generated speech. This phenomenon also accounts for the effectiveness of the Sway Sampling strategy proposed in F5-TTS \cite{chen2024f5}, which allocates more flow steps for the early stage of reverse generation.

\begin{figure*} [h]
    \begin{minipage}{0.99\linewidth}
        \centerline{\includegraphics[width=\textwidth]{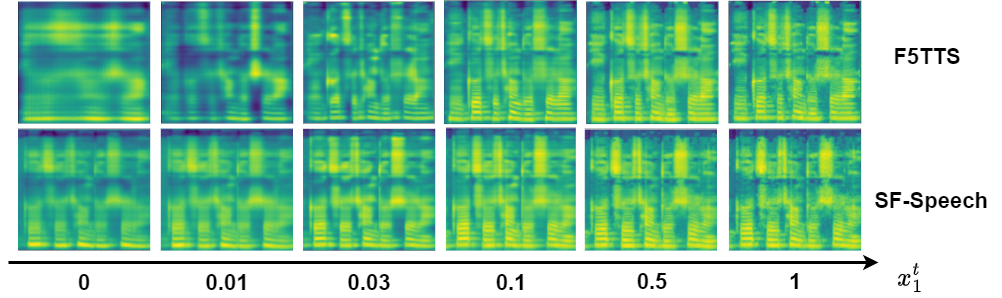}}
    \end{minipage}
\centering
\caption{Extrapolation of the terminal values $\hat{x^t_1}$ at different reverse time step $t$ for SF-Speech and F5-TTS.}
\label{fig8}
\end{figure*}

Furthermore, to further visualize the trajectory curvature, we extrapolate the terminal values at reverse time step $t$, $\hat{x^t_1} = x_t + (1-t)v(x_t)$, as illustrated in Fig. \ref{fig7}. Fig. \ref{fig8} shows the $\hat{x^t_1}$ of SF-Speech and F5-TTS. Compared with F5-TTS, the terminal extrapolations of SF-Speech at different time steps are more stable. This further indicates that the straightened flow $v_t$ in SF-Speech points directly to the desirable result in the early stage of reverse generation. However, it is worth noting that reducing curvature only serves to stabilize the direction of the generation trajectory, thus allowing high-quality speech synthesis with a small NFEs. It is unable to optimize the terminal value $\hat{x}_1$, which in turn determines the upper bound of the generated speech quality. Whether the $\hat{x}_1$ follows the real target distribution hinges on the complexity of tasks and the learning capacity of networks. Fortunately, as we analyzed in Sec. \ref{OMFM}, learning the mapping between deterministically coupled distributions is considerably more straightforward than learning that between arbitrarily coupled ones. The superiority of SF-Speech in generated speech quality demonstrates this phenomenon compellingly.


\begin{table}[htbp]
  \begin{center}
  \caption{Ablation studies on network structure. V1 represents the model using version 1 of detail ODE, while V2 represents version 2 of detail ODE.}
  \label{tab4}
  \begin{tabular}{c|c|c c} 
  \toprule [2pt]
   \textbf{Task}& \textbf{Model} & \textbf{SIM-o} & \textbf{WER}\\
  \midrule 
  \multirow{2}{*}{ZS-TTS}&
    Proposed V1           &\textbf{0.545}&12.41\\
    &Proposed V2           &0.543&\textbf{11.46}\\
  \midrule 
  \multirow{2}{*}{ZS-SR}&
    Proposed V1            &0.692&8.69\\
    &Proposed V2   &\textbf{0.694}&\textbf{8.51}\\
  \bottomrule [2pt]
  \end{tabular}
  \end{center}
\end{table}

\subsection{Network Comparison}
\label{nc}
Within this section, we evaluated the effectiveness of 2D convolution on the tasks ZS-TTS and ZS-SR. The compared results are shown in Tab \ref{tab3}. After employing 2D convolution for the fusion of 2D local information of input features on frequency and time axes, the proposed model shows a performance improvement on WER, with a decrease of 0.95\% on the ZS-TTS task and 0.18\% on the ZS-SR task. However, 2D convolution did not improve speaker similarity, with a SIM-o improvement of 0.002 on the ZS-SR task but a SIM-o decrease of 0.002 on the ZS-TTS task. We think this is because SIM is computed on global speaker embedding extracted from the whole audio, whereas 2D convolution can only assist the proposed model in modeling local information. While the 2D CNN cannot assist the model in speaker similarity, its contribution to speech intelligibility demonstrates that it is helpful for modeling features with high local correlation on time and channel axes. This result implies that designing networks according to the space correlation for different acoustic features is one of the ways to get the most out of these features.

\begin{table}[htbp]
  \begin{center}
  \caption{Ablation studies on training strategy and condition features. F1 and F2 represent the Text Feature and Coarse Feature. M means training masked frames only.}
  \label{tab5}
  \begin{tabular}{c|c c} 
  \toprule [2pt]
   \textbf{Condition and Strategy} & \textbf{SIM-o} & \textbf{WER}\\
  \midrule 
        F1+F2+M(proposed) &0.545&12.41\\
        F1+F2             &0.531&16.13\\
        F2                &0.533&17.94\\
        F1                &0.525&17.51\\
  \bottomrule [2pt]
  \end{tabular}
  \end{center}
\end{table}

\subsection{Ablation Study}

In this section, we first tested the effect of focusing on the masked speech. Then, we ablated the conditional features of the detail ODE module to verify that the different features contained different information, as we hypothesized in Sec \ref{OM}. We trained the SF-Speech containing version 1 of detail ODE with different training strategies and ODE conditional features to test their performance on the ZS-TTS task. As shown in Tab \ref{tab4}, training only the masked frames significantly improves the timbre similarity and intelligibility of generated speech, with an improvement of 0.014 on SIM-o and a decrease of 3.72\% on WER. This result suggests that computing only the loss of masked frames is crucial for these ODE-based models with in-context masking patterns, which is consistent with the previous work \cite{le2023voicebox}. In addition, ablating different conditional features leads to different variations in the SIM-o and WER. When the F1 is removed, the SIM-o increases by 0.002, while the WER increases by 1.81\%. This result implies that F1  contains only pronunciation information and text-related prosody. When the F2 is removed, the SIM-o of the generated speech decreases by 0.006, and the WER increases by 1.38\%, suggesting our hypothesis in Sec. \ref{OM} that F2 combines speaker information, the speaker-related prosody, and F1.

\section{Conclusion}
\label{cl}
In this paper, we proposed SF-Speech, a novel ODE-based method for zero-shot voice clone. Specifically, SF-Speech employs a new two-stage method to generate representations that correspond to speech unambiguously in mel-spectrogram space and utilizes this deterministic coupling pair to straighten the reverse trajectories learned by the ODE model. We conducted a large number of experiments on datasets of diverse scales. In comparison with other large-scale TTS models, SF-Speech achieves a new SOTA performance with faster inference speed. Moreover, owing to the design of the initial distribution, SF-Speech can maintain outstanding performance when dealing with small-scale datasets and in-the-wild data. 
To further understand the underlying mechanisms, we carried out in-depth analyses of the reverse trajectory curvature in TTS models based on FM. Empirically, we demonstrate that the straightened flow significantly enhances both the generation efficiency and stability. Additionally, we introduced a simple yet effective 2D convolution plugin in the Transformer backbone of the ODE model, further enhancing the quality of the generated mel spectrogram.

However, we also encountered some challenges. We find that voice cloning relying on in-context learning is highly sensitive to the quality of the training data. The audio imperfections present in the wild data lead to a degradation in the quality of the generated speech, which becomes more pronounced as the number of ODE function evaluations increases. We surmise this is because these models capture not only speaker information but also audio imperfections within the audio context.  In the future, our research will focus on exploring the decoupling of speaker information and audio imperfections from the audio context. 


\bibliographystyle{IEEEtran}
\bibliography{reference}

\begin{thebibliography}{10}
\providecommand{\url}[1]{#1}
\csname url@samestyle\endcsname
\providecommand{\newblock}{\relax}
\providecommand{\bibinfo}[2]{#2}
\providecommand{\BIBentrySTDinterwordspacing}{\spaceskip=0pt\relax}
\providecommand{\BIBentryALTinterwordstretchfactor}{4}
\providecommand{\BIBentryALTinterwordspacing}{\spaceskip=\fontdimen2\font plus
\BIBentryALTinterwordstretchfactor\fontdimen3\font minus \fontdimen4\font\relax}
\providecommand{\BIBforeignlanguage}[2]{{%
\expandafter\ifx\csname l@#1\endcsname\relax
\typeout{** WARNING: IEEEtran.bst: No hyphenation pattern has been}%
\typeout{** loaded for the language `#1'. Using the pattern for}%
\typeout{** the default language instead.}%
\else
\language=\csname l@#1\endcsname
\fi
#2}}
\providecommand{\BIBdecl}{\relax}
\BIBdecl

\bibitem{gibiansky2017deep}
A.~Gibiansky, S.~Arik, G.~Diamos, J.~Miller, K.~Peng, W.~Ping, J.~Raiman, and Y.~Zhou, ``Deep voice 2: Multi-speaker neural text-to-speech,'' \emph{Advances in neural information processing systems}, vol.~30, 2017.

\bibitem{moss2020boffin}
H.~B. Moss, V.~Aggarwal, N.~Prateek, J.~Gonz{\'a}lez, and R.~Barra-Chicote, ``Boffin tts: Few-shot speaker adaptation by bayesian optimization,'' in \emph{ICASSP 2020-2020 IEEE International Conference on Acoustics, Speech and Signal Processing (ICASSP)}.\hskip 1em plus 0.5em minus 0.4em\relax IEEE, 2020, pp. 7639--7643.

\bibitem{huang2022meta}
S.-F. Huang, C.-J. Lin, D.-R. Liu, Y.-C. Chen, and H.-y. Lee, ``Meta-tts: Meta-learning for few-shot speaker adaptive text-to-speech,'' \emph{IEEE/ACM Transactions on Audio, Speech, and Language Processing}, vol.~30, pp. 1558--1571, 2022.

\bibitem{min2021meta}
D.~Min, D.~B. Lee, E.~Yang, and S.~J. Hwang, ``Meta-stylespeech: Multi-speaker adaptive text-to-speech generation,'' in \emph{International Conference on Machine Learning}.\hskip 1em plus 0.5em minus 0.4em\relax PMLR, 2021, pp. 7748--7759.

\bibitem{jiang2023mega}
Z.~Jiang, Y.~Ren, Z.~Ye, J.~Liu, C.~Zhang, Q.~Yang, S.~Ji, R.~Huang, C.~Wang, X.~Yin \emph{et~al.}, ``Mega-tts: Zero-shot text-to-speech at scale with intrinsic inductive bias,'' \emph{arXiv preprint arXiv:2306.03509}, 2023.

\bibitem{casanova2022yourtts}
E.~Casanova, J.~Weber, C.~D. Shulby, A.~C. Junior, E.~G{\"o}lge, and M.~A. Ponti, ``Yourtts: Towards zero-shot multi-speaker tts and zero-shot voice conversion for everyone,'' in \emph{International Conference on Machine Learning}.\hskip 1em plus 0.5em minus 0.4em\relax PMLR, 2022, pp. 2709--2720.

\bibitem{li2023freevc}
J.~Li, W.~Tu, and L.~Xiao, ``Freevc: Towards high-quality text-free one-shot voice conversion,'' in \emph{ICASSP 2023-2023 IEEE International Conference on Acoustics, Speech and Signal Processing (ICASSP)}.\hskip 1em plus 0.5em minus 0.4em\relax IEEE, 2023, pp. 1--5.

\bibitem{wang2023neural}
C.~Wang, S.~Chen, Y.~Wu, Z.~Zhang, L.~Zhou, S.~Liu, Z.~Chen, Y.~Liu, H.~Wang, J.~Li \emph{et~al.}, ``Neural codec language models are zero-shot text to speech synthesizers,'' \emph{arXiv preprint arXiv:2301.02111}, 2023.

\bibitem{yang2023uniaudio}
D.~Yang, J.~Tian, X.~Tan, R.~Huang, S.~Liu, X.~Chang, J.~Shi, S.~Zhao, J.~Bian, X.~Wu \emph{et~al.}, ``Uniaudio: An audio foundation model toward universal audio generation,'' \emph{arXiv preprint arXiv:2310.00704}, 2023.

\bibitem{casanova2024xtts}
E.~Casanova, K.~Davis, E.~G{\"o}lge, G.~G{\"o}knar, I.~Gulea, L.~Hart, A.~Aljafari, J.~Meyer, R.~Morais, S.~Olayemi \emph{et~al.}, ``Xtts: a massively multilingual zero-shot text-to-speech model,'' \emph{arXiv e-prints}, pp. arXiv--2406, 2024.

\bibitem{floridi2020gpt}
L.~Floridi and M.~Chiriatti, ``Gpt-3: Its nature, scope, limits, and consequences,'' \emph{Minds and Machines}, vol.~30, pp. 681--694, 2020.

\bibitem{touvron2023llama}
H.~Touvron, T.~Lavril, G.~Izacard, X.~Martinet, M.-A. Lachaux, T.~Lacroix, B.~Rozi{\`e}re, N.~Goyal, E.~Hambro, F.~Azhar \emph{et~al.}, ``Llama: Open and efficient foundation language models,'' \emph{arXiv preprint arXiv:2302.13971}, 2023.

\bibitem{zeghidour2021soundstream}
N.~Zeghidour, A.~Luebs, A.~Omran, J.~Skoglund, and M.~Tagliasacchi, ``Soundstream: An end-to-end neural audio codec,'' \emph{IEEE/ACM Transactions on Audio, Speech, and Language Processing}, vol.~30, pp. 495--507, 2021.

\bibitem{defossez2022high}
A.~D{\'e}fossez, J.~Copet, G.~Synnaeve, and Y.~Adi, ``High fidelity neural audio compression,'' \emph{arXiv preprint arXiv:2210.13438}, 2022.

\bibitem{kumar2024high}
R.~Kumar, P.~Seetharaman, A.~Luebs, I.~Kumar, and K.~Kumar, ``High-fidelity audio compression with improved rvqgan,'' \emph{Advances in Neural Information Processing Systems}, vol.~36, 2024.

\bibitem{ju2024naturalspeech}
Z.~Ju, Y.~Wang, K.~Shen, X.~Tan, D.~Xin, D.~Yang, Y.~Liu, Y.~Leng, K.~Song, S.~Tang \emph{et~al.}, ``Naturalspeech 3: Zero-shot speech synthesis with factorized codec and diffusion models,'' \emph{arXiv preprint arXiv:2403.03100}, 2024.

\bibitem{austin2021structured}
J.~Austin, D.~D. Johnson, J.~Ho, D.~Tarlow, and R.~Van Den~Berg, ``Structured denoising diffusion models in discrete state-spaces,'' \emph{Advances in Neural Information Processing Systems}, vol.~34, pp. 17\,981--17\,993, 2021.

\bibitem{lajszczak2024base}
M.~{\L}ajszczak, G.~C{\'a}mbara, Y.~Li, F.~Beyhan, A.~van Korlaar, F.~Yang, A.~Joly, {\'A}.~Mart{\'\i}n-Cortinas, A.~Abbas, A.~Michalski \emph{et~al.}, ``Base tts: Lessons from building a billion-parameter text-to-speech model on 100k hours of data,'' \emph{arXiv preprint arXiv:2402.08093}, 2024.

\bibitem{meng2024autoregressive}
L.~Meng, L.~Zhou, S.~Liu, S.~Chen, B.~Han, S.~Hu, Y.~Liu, J.~Li, S.~Zhao, X.~Wu \emph{et~al.}, ``Autoregressive speech synthesis without vector quantization,'' \emph{arXiv preprint arXiv:2407.08551}, 2024.

\bibitem{wang2024evaluating}
S.~Wang and {\'E}.~Sz{\'e}kely, ``Evaluating text-to-speech synthesis from a large discrete token-based speech language model,'' \emph{arXiv preprint arXiv:2405.09768}, 2024.

\bibitem{qiang2024high}
C.~Qiang, H.~Li, Y.~Tian, Y.~Zhao, Y.~Zhang, L.~Wang, and J.~Dang, ``High-fidelity speech synthesis with minimal supervision: All using diffusion models,'' in \emph{ICASSP 2024-2024 IEEE International Conference on Acoustics, Speech and Signal Processing (ICASSP)}.\hskip 1em plus 0.5em minus 0.4em\relax IEEE, 2024, pp. 10\,781--10\,785.

\bibitem{shen2023naturalspeech}
K.~Shen, Z.~Ju, X.~Tan, Y.~Liu, Y.~Leng, L.~He, T.~Qin, S.~Zhao, and J.~Bian, ``Naturalspeech 2: Latent diffusion models are natural and zero-shot speech and singing synthesizers,'' \emph{arXiv preprint arXiv:2304.09116}, 2023.

\bibitem{le2023voicebox}
M.~Le, A.~Vyas, B.~Shi, B.~Karrer, L.~Sari, R.~Moritz, M.~Williamson, V.~Manohar, Y.~Adi, J.~Mahadeokar \emph{et~al.}, ``Voicebox: Text-guided multilingual universal speech generation at scale,'' \emph{arXiv preprint arXiv:2306.15687}, 2023.

\bibitem{kim2024p}
S.~Kim, K.~Shih, J.~F. Santos, E.~Bakhturina, M.~Desta, R.~Valle, S.~Yoon, B.~Catanzaro \emph{et~al.}, ``P-flow: A fast and data-efficient zero-shot tts through speech prompting,'' \emph{Advances in Neural Information Processing Systems}, vol.~36, 2024.

\bibitem{chen2018neural}
R.~T. Chen, Y.~Rubanova, J.~Bettencourt, and D.~K. Duvenaud, ``Neural ordinary differential equations,'' \emph{Advances in neural information processing systems}, vol.~31, 2018.

\bibitem{lipman2022flow}
Y.~Lipman, R.~T. Chen, H.~Ben-Hamu, M.~Nickel, and M.~Le, ``Flow matching for generative modeling,'' \emph{arXiv preprint arXiv:2210.02747}, 2022.

\bibitem{eskimez2024e2}
S.~E. Eskimez, X.~Wang, M.~Thakker, C.~Li, C.-H. Tsai, Z.~Xiao, H.~Yang, Z.~Zhu, M.~Tang, X.~Tan \emph{et~al.}, ``E2 tts: Embarrassingly easy fully non-autoregressive zero-shot tts,'' in \emph{2024 IEEE Spoken Language Technology Workshop (SLT)}.\hskip 1em plus 0.5em minus 0.4em\relax IEEE, 2024, pp. 682--689.

\bibitem{chen2024f5}
Y.~Chen, Z.~Niu, Z.~Ma, K.~Deng, C.~Wang, J.~Zhao, K.~Yu, and X.~Chen, ``F5-tts: A fairytaler that fakes fluent and faithful speech with flow matching,'' \emph{arXiv preprint arXiv:2410.06885}, 2024.

\bibitem{yang2024simplespeech}
D.~Yang, R.~Huang, Y.~Wang, H.~Guo, D.~Chong, S.~Liu, X.~Wu, and H.~Meng, ``Simplespeech 2: Towards simple and efficient text-to-speech with flow-based scalar latent transformer diffusion models,'' \emph{arXiv preprint arXiv:2408.13893}, 2024.

\bibitem{lee2023minimizing}
S.~Lee, B.~Kim, and J.~C. Ye, ``Minimizing trajectory curvature of ode-based generative models,'' in \emph{International Conference on Machine Learning}.\hskip 1em plus 0.5em minus 0.4em\relax PMLR, 2023, pp. 18\,957--18\,973.

\bibitem{liu2022flow}
X.~Liu, C.~Gong, and Q.~Liu, ``Flow straight and fast: Learning to generate and transfer data with rectified flow,'' \emph{arXiv preprint arXiv:2209.03003}, 2022.

\bibitem{song2020score}
Y.~Song, J.~Sohl-Dickstein, D.~P. Kingma, A.~Kumar, S.~Ermon, and B.~Poole, ``Score-based generative modeling through stochastic differential equations,'' \emph{arXiv preprint arXiv:2011.13456}, 2020.

\bibitem{hyvarinen2005estimation}
A.~Hyv{\"a}rinen and P.~Dayan, ``Estimation of non-normalized statistical models by score matching.'' \emph{Journal of Machine Learning Research}, vol.~6, no.~4, 2005.

\bibitem{rombach2022high}
R.~Rombach, A.~Blattmann, D.~Lorenz, P.~Esser, and B.~Ommer, ``High-resolution image synthesis with latent diffusion models,'' in \emph{Proceedings of the IEEE/CVF conference on computer vision and pattern recognition}, 2022, pp. 10\,684--10\,695.

\bibitem{ramesh2022hierarchical}
A.~Ramesh, P.~Dhariwal, A.~Nichol, C.~Chu, and M.~Chen, ``Hierarchical text-conditional image generation with clip latents,'' \emph{arXiv preprint arXiv:2204.06125}, vol.~1, no.~2, p.~3, 2022.

\bibitem{kong2020diffwave}
Z.~Kong, W.~Ping, J.~Huang, K.~Zhao, and B.~Catanzaro, ``Diffwave: A versatile diffusion model for audio synthesis,'' \emph{arXiv preprint arXiv:2009.09761}, 2020.

\bibitem{chen2020wavegrad}
N.~Chen, Y.~Zhang, H.~Zen, R.~J. Weiss, M.~Norouzi, and W.~Chan, ``Wavegrad: Estimating gradients for waveform generation,'' \emph{arXiv preprint arXiv:2009.00713}, 2020.

\bibitem{koizumi2022specgrad}
Y.~Koizumi, H.~Zen, K.~Yatabe, N.~Chen, and M.~Bacchiani, ``Specgrad: Diffusion probabilistic model based neural vocoder with adaptive noise spectral shaping,'' \emph{arXiv preprint arXiv:2203.16749}, 2022.

\bibitem{popov2021grad}
V.~Popov, I.~Vovk, V.~Gogoryan, T.~Sadekova, and M.~Kudinov, ``Grad-tts: A diffusion probabilistic model for text-to-speech,'' in \emph{International Conference on Machine Learning}.\hskip 1em plus 0.5em minus 0.4em\relax PMLR, 2021, pp. 8599--8608.

\bibitem{jing2023u}
X.~Jing, Y.~Chang, Z.~Yang, J.~Xie, A.~Triantafyllopoulos, and B.~W. Schuller, ``U-dit tts: U-diffusion vision transformer for text-to-speech,'' in \emph{Speech Communication; 15th ITG Conference}.\hskip 1em plus 0.5em minus 0.4em\relax VDE, 2023, pp. 56--60.

\bibitem{zheng2022truncated}
H.~Zheng, P.~He, W.~Chen, and M.~Zhou, ``Truncated diffusion probabilistic models,'' \emph{arXiv preprint arXiv:2202.09671}, vol.~1, no.~3, pp. 1--2, 2022.

\bibitem{lyu2022accelerating}
Z.~Lyu, X.~Xu, C.~Yang, D.~Lin, and B.~Dai, ``Accelerating diffusion models via early stop of the diffusion process,'' \emph{arXiv preprint arXiv:2205.12524}, 2022.

\bibitem{kong2021fast}
Z.~Kong and W.~Ping, ``On fast sampling of diffusion probabilistic models,'' \emph{arXiv preprint arXiv:2106.00132}, 2021.

\bibitem{lu2022dpm}
C.~Lu, Y.~Zhou, F.~Bao, J.~Chen, C.~Li, and J.~Zhu, ``Dpm-solver: A fast ode solver for diffusion probabilistic model sampling in around 10 steps,'' \emph{Advances in Neural Information Processing Systems}, vol.~35, pp. 5775--5787, 2022.

\bibitem{sun2017ito}
Y.~Sun, J.~Yang, and W.~Zhao, ``Ito-taylor schemes for solving mean-field stochastic differential equations,'' \emph{Numerical Mathematics: Theory, Methods and Applications}, vol.~10, no.~4, pp. 798--828, 2017.

\bibitem{mehta2024matcha}
S.~Mehta, R.~Tu, J.~Beskow, {\'E}.~Sz{\'e}kely, and G.~E. Henter, ``Matcha-tts: A fast tts architecture with conditional flow matching,'' in \emph{ICASSP 2024-2024 IEEE International Conference on Acoustics, Speech and Signal Processing (ICASSP)}.\hskip 1em plus 0.5em minus 0.4em\relax IEEE, 2024, pp. 11\,341--11\,345.

\bibitem{guo2024voiceflow}
Y.~Guo, C.~Du, Z.~Ma, X.~Chen, and K.~Yu, ``Voiceflow: Efficient text-to-speech with rectified flow matching,'' in \emph{ICASSP 2024-2024 IEEE International Conference on Acoustics, Speech and Signal Processing (ICASSP)}.\hskip 1em plus 0.5em minus 0.4em\relax IEEE, 2024, pp. 11\,121--11\,125.

\bibitem{higgins2016beta}
I.~Higgins, L.~Matthey, A.~Pal, C.~Burgess, X.~Glorot, M.~Botvinick, S.~Mohamed, and A.~Lerchner, ``beta-vae: Learning basic visual concepts with a constrained variational framework,'' in \emph{International conference on learning representations}, 2016.

\bibitem{chen2024vall}
S.~Chen, S.~Liu, L.~Zhou, Y.~Liu, X.~Tan, J.~Li, S.~Zhao, Y.~Qian, and F.~Wei, ``Vall-e 2: Neural codec language models are human parity zero-shot text to speech synthesizers,'' \emph{arXiv preprint arXiv:2406.05370}, 2024.

\bibitem{anastassiou2024seed}
P.~Anastassiou, J.~Chen, J.~Chen, Y.~Chen, Z.~Chen, Z.~Chen, J.~Cong, L.~Deng, C.~Ding, L.~Gao \emph{et~al.}, ``Seed-tts: A family of high-quality versatile speech generation models,'' \emph{arXiv preprint arXiv:2406.02430}, 2024.

\bibitem{du2024cosyvoice}
Z.~Du, Q.~Chen, S.~Zhang, K.~Hu, H.~Lu, Y.~Yang, H.~Hu, S.~Zheng, Y.~Gu, Z.~Ma \emph{et~al.}, ``Cosyvoice: A scalable multilingual zero-shot text-to-speech synthesizer based on supervised semantic tokens,'' \emph{arXiv preprint arXiv:2407.05407}, 2024.

\bibitem{du2024cosyvoice2}
Z.~Du, Y.~Wang, Q.~Chen, X.~Shi, X.~Lv, T.~Zhao, Z.~Gao, Y.~Yang, C.~Gao, H.~Wang \emph{et~al.}, ``Cosyvoice 2: Scalable streaming speech synthesis with large language models,'' \emph{arXiv preprint arXiv:2412.10117}, 2024.

\bibitem{li2022styletts}
Y.~A. Li, C.~Han, and N.~Mesgarani, ``Styletts: A style-based generative model for natural and diverse text-to-speech synthesis,'' \emph{arXiv preprint arXiv:2205.15439}, 2022.

\bibitem{ren2019fastspeech}
Y.~Ren, Y.~Ruan, X.~Tan, T.~Qin, S.~Zhao, Z.~Zhao, and T.-Y. Liu, ``Fastspeech: Fast, robust and controllable text to speech,'' \emph{Advances in neural information processing systems}, vol.~32, 2019.

\bibitem{kim2021conditional}
J.~Kim, J.~Kong, and J.~Son, ``Conditional variational autoencoder with adversarial learning for end-to-end text-to-speech,'' in \emph{International Conference on Machine Learning}.\hskip 1em plus 0.5em minus 0.4em\relax PMLR, 2021, pp. 5530--5540.

\bibitem{yang2023hifi}
D.~Yang, S.~Liu, R.~Huang, J.~Tian, C.~Weng, and Y.~Zou, ``Hifi-codec: Group-residual vector quantization for high fidelity audio codec,'' \emph{arXiv preprint arXiv:2305.02765}, 2023.

\bibitem{peebles2023scalable}
W.~Peebles and S.~Xie, ``Scalable diffusion models with transformers,'' in \emph{Proceedings of the IEEE/CVF International Conference on Computer Vision}, 2023, pp. 4195--4205.

\bibitem{bao2023all}
F.~Bao, S.~Nie, K.~Xue, Y.~Cao, C.~Li, H.~Su, and J.~Zhu, ``All are worth words: A vit backbone for diffusion models,'' in \emph{Proceedings of the IEEE/CVF conference on computer vision and pattern recognition}, 2023, pp. 22\,669--22\,679.

\bibitem{2017CSTR}
C.~Veaux, J.~Yamagishi, and K.~MacDonald, ``Cstr vctk corpus: English multi-speaker corpus for cstr voice cloning toolkit,'' 2017.

\bibitem{panayotov2015librispeech}
V.~Panayotov, G.~Chen, D.~Povey, and S.~Khudanpur, ``Librispeech: an asr corpus based on public domain audio books,'' in \emph{2015 IEEE international conference on acoustics, speech and signal processing (ICASSP)}.\hskip 1em plus 0.5em minus 0.4em\relax IEEE, 2015, pp. 5206--5210.

\bibitem{ljspeech17}
K.~Ito and L.~Johnson, ``The lj speech dataset,'' \url{https://keithito.com/LJ-Speech-Dataset/}, 2017.

\bibitem{nguyen2023expresso}
T.~A. Nguyen, W.-N. Hsu, A.~d'Avirro, B.~Shi, I.~Gat, M.~Fazel-Zarani, T.~Remez, J.~Copet, G.~Synnaeve, M.~Hassid \emph{et~al.}, ``Expresso: A benchmark and analysis of discrete expressive speech resynthesis,'' \emph{arXiv preprint arXiv:2308.05725}, 2023.

\bibitem{csmsc17}
D.~Baker, ``Chinese standard mandarin speech corpus,'' \url{https://www.data-baker.com/open source.html}, 2017.

\bibitem{shi2020aishell}
Y.~Shi, H.~Bu, X.~Xu, S.~Zhang, and M.~Li, ``{AISHELL-3: A Multi-Speaker Mandarin TTS Corpus},'' in \emph{Proc. Interspeech 2021}, 2021, pp. 2756--2760.

\bibitem{gong2022ssast}
Y.~Gong, C.-I. Lai, Y.-A. Chung, and J.~Glass, ``Ssast: Self-supervised audio spectrogram transformer,'' in \emph{Proceedings of the AAAI Conference on Artificial Intelligence}, vol.~36, no.~10, 2022, pp. 10\,699--10\,709.

\bibitem{guo2022multi}
H.~Guo, H.~Lu, X.~Wu, and H.~Meng, ``A multi-scale time-frequency spectrogram discriminator for gan-based non-autoregressive tts,'' \emph{arXiv preprint arXiv:2203.01080}, 2022.

\bibitem{gulati2020conformer}
A.~Gulati, J.~Qin, C.-C. Chiu, N.~Parmar, Y.~Zhang, J.~Yu, W.~Han, S.~Wang, Z.~Zhang, Y.~Wu \emph{et~al.}, ``Conformer: Convolution-augmented transformer for speech recognition,'' \emph{arXiv preprint arXiv:2005.08100}, 2020.

\bibitem{he2024emilia}
H.~He, Z.~Shang, C.~Wang, X.~Li, Y.~Gu, H.~Hua, L.~Liu, C.~Yang, J.~Li, P.~Shi \emph{et~al.}, ``Emilia: An extensive, multilingual, and diverse speech dataset for large-scale speech generation,'' in \emph{2024 IEEE Spoken Language Technology Workshop (SLT)}.\hskip 1em plus 0.5em minus 0.4em\relax IEEE, 2024, pp. 885--890.

\bibitem{magic19}
L.~Magic Data Technology~Co., ``Magicdata mandarin chinese read speech corpus,'' \url{https://www.openslr.org/68/}, 2019.

\bibitem{Bernard2021}
\BIBentryALTinterwordspacing
M.~Bernard and H.~Titeux, ``Phonemizer: Text to phones transcription for multiple languages in python,'' \emph{Journal of Open Source Software}, vol.~6, no.~68, p. 3958, 2021. [Online]. Available: \url{https://doi.org/10.21105/joss.03958}
\BIBentrySTDinterwordspacing

\bibitem{2017Montreal}
M.~Mcauliffe, M.~Socolof, S.~Mihuc, M.~Wagner, and M.~Sonderegger, ``Montreal forced aligner: Trainable text-speech alignment using kaldi,'' in \emph{Interspeech}, 2017, pp. 498--502.

\bibitem{lee2022bigvgan}
S.-g. Lee, W.~Ping, B.~Ginsburg, B.~Catanzaro, and S.~Yoon, ``Bigvgan: A universal neural vocoder with large-scale training,'' \emph{arXiv preprint arXiv:2206.04658}, 2022.

\bibitem{siuzdak2023vocos}
H.~Siuzdak, ``Vocos: Closing the gap between time-domain and fourier-based neural vocoders for high-quality audio synthesis,'' \emph{arXiv preprint arXiv:2306.00814}, 2023.

\bibitem{ulyanov2016instance}
D.~Ulyanov, A.~Vedaldi, and V.~Lempitsky, ``Instance normalization: The missing ingredient for fast stylization,'' \emph{arXiv preprint arXiv:1607.08022}, 2016.

\bibitem{ho2022classifier}
J.~Ho and T.~Salimans, ``Classifier-free diffusion guidance,'' \emph{arXiv preprint arXiv:2207.12598}, 2022.

\bibitem{heo2020clova}
H.~S. Heo, B.-J. Lee, J.~Huh, and J.~S. Chung, ``Clova baseline system for the voxceleb speaker recognition challenge 2020,'' \emph{arXiv preprint arXiv:2009.14153}, 2020.

\bibitem{zhang2022wenet}
B.~Zhang, D.~Wu, Z.~Peng, X.~Song, Z.~Yao, H.~Lv, L.~Xie, C.~Yang, F.~Pan, and J.~Niu, ``Wenet 2.0: More productive end-to-end speech recognition toolkit,'' \emph{arXiv preprint arXiv:2203.15455}, 2022.

\bibitem{reddy2022dnsmos}
C.~K. Reddy, V.~Gopal, and R.~Cutler, ``Dnsmos p. 835: A non-intrusive perceptual objective speech quality metric to evaluate noise suppressors,'' in \emph{ICASSP 2022-2022 IEEE International Conference on Acoustics, Speech and Signal Processing (ICASSP)}.\hskip 1em plus 0.5em minus 0.4em\relax IEEE, 2022, pp. 886--890.

\bibitem{desplanques2020ecapa}
B.~Desplanques, J.~Thienpondt, and K.~Demuynck, ``Ecapa-tdnn: Emphasized channel attention, propagation and aggregation in tdnn based speaker verification,'' \emph{arXiv preprint arXiv:2005.07143}, 2020.

\end{thebibliography}


 




\vfill

\end{document}